\documentclass[12pt]{article}

\usepackage{geometry}                % See geometry.pdf to learn the layout options. There are lots.
\usepackage{graphicx}
\usepackage{amssymb}
\usepackage{epstopdf}
\usepackage{slashed}
\usepackage{float}
\usepackage{hyperref}
\usepackage{amsmath}
\usepackage{caption}
\usepackage{subcaption}
\numberwithin{equation}{section}

\newcommand{\be}{\begin{equation}}
\newcommand{\ee}{\end{equation}}
\newcommand{\bea}{\begin{eqnarray}}
\newcommand{\eea}{\end{eqnarray}}
\newcommand{\beas}{\begin{eqnarray*}}
\newcommand{\eeas}{\end{eqnarray*}}
\newcommand{\ba}{\begin{array}}
\newcommand{\ea}{\end{array}}

\newcommand{\nn}{\nonumber}

\def\be{\begin{equation}}
\def\ee{\end{equation}}
\def\ben{\begin{equation*}}
\def\een{\end{equation*}}
\def\beqa{\begin{eqnarray}}
\def\eeqa{\end{eqnarray}}

\newcommand{\tlr}{\tilde{r}}

\renewcommand{\d}{\partial}

\renewcommand{\(}{\left(}
\renewcommand{\)}{\right)}
\renewcommand{\[}{\left[}
\renewcommand{\]}{\right]}

\newcommand{\half}{\frac{1}{2}}
\newcommand{\rootg}{\sqrt{-g}}
\newcommand{\dr}{\partial_r}
\newcommand{\dx}{\partial_x}
\newcommand{\drho}{\partial_\rho}
\newcommand{\ohat}{\hat{O}}
\newcommand{\tilr}{\tilde{r}}
\newcommand{\tilx}{\tilde{x}}
\newcommand{\rhoi}{\rho_{int}}

\def\bi{\begin{itemize}}
\def\ei{\end{itemize}}

\title{\bf Striped Order in AdS/CFT}
\date{\today}
\author{ Moshe Rozali, Darren Smyth, Evgeny Sorkin and Jared B. Stang \\ \\
\small Department of Physics and Astronomy,\\
\small University of British Columbia,\\
\small Vancouver, BC V6T 1Z1, Canada\\
\small \texttt{rozali, dsmyth, evgeny, jstang~\, @phas.ubc.ca} }

\begin{document}
\maketitle

%====================================================================%
\begin{abstract}
We study the formation of inhomogeneous order in the Einstein-Maxwell-axion system, dual to a 2+1 dimensional field theory that exhibits a spontaneously generated current density, momentum density and modulated scalar operator. Below the critical temperature, the Reissner-Nordstr\"{o}m-AdS black hole becomes unstable and stripes form in the bulk and on the boundary. The bulk geometry possesses striking geometrical features, including a modulated horizon that tends to pinch off as $T\rightarrow0$. On a domain of fixed length, we find a second order phase transition to the striped solution in each of the grand canonical, canonical and microcanonical ensembles, with modulated charges that grow and saturate as we lower the temperature and descend into the inhomogeneous phase. For the black hole on an infinite domain, a similar second order transition occurs, and the width of the dominant stripe increases in the zero temperature limit.
\end{abstract}
%====================================================================%

%====================================================================%
\newpage
\section{Introduction and Summary}
 
The gauge-gravity duality is a relationship between a strongly coupled field theory and a gravity system in one higher dimension. This correspondence has been fruitful in studying various field theory phenomena by translating the problem to the gravitational context. In particular, the duality has shone new light on many condensed matter systems - see \cite{Hartnoll:2009sz,Herzog:2009xv,McGreevy:2009xe,Sachdev:2011wg} for reviews.

Early models in this area, such as the holographic superconductor \cite{Hartnoll:2008kx}, focused on homogeneous phases of  field theories. In this case, the fields on the gravity side depend only on the radial coordinate in the bulk and the problem reduces to the solution of ODEs. However, many interesting phenomena occur in less symmetric situations. Generically, the problem of finding the gravity dual to an inhomogeneous boundary system will necessitate solving relatively more difficult PDEs, almost always resulting in the need for numerical methods. While these become technically hard problems, there exist established numerical approaches. Due to the success of the holographic method in studying homogeneous situations, it is worthwhile to push the correspondence to these less symmetric situations in order to describe more general phenomena in this context.

One particular area of condensed matter that appears to be amenable to a holographic description is the appearance of striped phases in certain materials.\footnote{Stripes are also known to form in large $N$ QCD \cite{Deryagin:1992rw,Shuster:1999tn}.} These phases are characterized by the spontaneous breaking of translational invariance in the system. Examples include charge density waves and spin density waves in strongly correlated electron systems, where either the charge and/or the spin densities become spatially modulated (for a review see \cite{vojta}).  The formation of stripes is conjectured to be related to the mechanism of superconductivity in the cuprates \cite{concepts}. To approach this striking phenomenon from the holographic perspective, one would look for an asymptotically AdS gravity system which allows a spontaneous transition to a modulated phase.

Recently, several interesting spatially modulated holographic systems have been studied. One way to study stripes on the boundary is to source them by imposing spatial modulation in the non-normalizable modes of some fields, explicitly breaking the translation invariance, as in \cite{Flauger:2010tv,Hutasoit:2012ib}.\footnote{In a similar vein, more recently, lattice-deformed black branes have been of interest in studies of conductivity in holographic models \cite{Donos:2012ra,Horowitz:2012gs,Horowitz:2012ky,Horowitz:2013jaa}.} However, if one wishes to make contact with the context described above, it is important that the inhomogeneity emerges spontaneously rather than be introduced explicitly.

In some cases, the spatially modulated phase has an extra symmetry, allowing the situation to be posed as a co-homogeneity one problem on the gravity side. Examples include systems in which one of the translational Killing vectors is replaced by a helical Killing vector \cite{Nakamura:2009tf,Ooguri:2010kt,Ooguri:2010xs,Donos:2011ff,Donos:2012gg,Donos:2012wi}. 
More general inhomogeneous instabilities, in which one of the translation symmetries is fully broken, have been described in a phenomenological model \cite{Donos:2011bh} and in certain $\#ND=6$ brane systems \cite{Bergman:2011rf,Jokela:2012vn,Jokela:2012se}.\footnote{Other studies of inhomogeneity in the context of holography include \cite{Iizuka:2012pn,Rozali:2012gf,Bu:2012mq,Bao:2013fda}.}

In this work, we study the full non-linear co-homogeneity two striped solutions to the Einstein-Maxwell-axion model that stem from the normalizable, inhomogeneous modes of the Reissner-Nordstr\"{o}m-AdS solution detailed in \cite{Donos:2011bh}. In this model, below a critical temperature, stripes spontaneously form in the bulk and on the boundary. We study the properties of the stripes in both the fixed length system, in which the wavenumber is set by the size of the domain and charges are integrated over the stripe, and the infinite system, in which the corresponding thermodynamic  densities are studied. For the black hole at fixed length, we examine the behavior in different thermodynamic ensembles as we vary the temperature and wavenumber. 

The study is facilitated by a numerical solution to the set of coupled Einstein and matter equations in the bulk. Inspired by the black string case \cite{Wiseman:2002zc,Sorkin:2006wp}, we fix the metric in the conformal gauge, resulting in a set of field equations and a set of constraint equations. Then, as described in \cite{Wiseman:2002zc}, the resulting constraint equations can be solved by imposing particular boundary conditions on the fields.

As well as being of interest from the holographic perspective these numerical solutions are important as they represent new inhomogeneous black hole solutions in Ads. We find strong evidence that the unstable homogeneous branes transition smoothly to the striped state below the critical temperature.\footnote{
The instability to the formation of the striped black branes resembles
the black string instability \cite{GL} which is known to be of the second order for high enough dimensions \cite{CritDim,Barak-Evgeny}.} 
As we approach zero temperature the relative inhomogeneity is seen to grow without bound and the black hole horizon tends to pinch off, signalling the formation of a spacetime singularity in this limit.

A subset of our results has already been announced in \cite{Stripes1}, in this paper we provide full details. The summary of the 
results follow:  
\subsubsection*{Boundary field theory}

\begin{itemize}
 \item We calculate the fully back-reacted normalizable inhomogeneous modes.

\item The stripes have momentum, electric current and modulations in charge and mass density (see \cite{Liu:2012zm} for a recent study of angular momentum generation).

\item As a function of temperature, the modulations start small, then grow and saturate as $T\rightarrow0$.

\item We study the stripe of fixed length in various ensembles, finding a second order phase transition, for sufficiently large axion coupling, in each of the grand canonical (temperature $T$, chemical potential $\mu$ fixed), canonical ($T$, charge $N$ fixed) and microcanonical (mass $M$, $N$ fixed) ensembles. We compute corresponding critical exponents. 

\item For the infinite length system, there is a second order transition to a striped phase. The width of the dominant stripe grows as the temperature is decreased.

\item In the zero temperature limit, within the accuracy of our numerics, the entropy appears to approach a non-zero value.

\end{itemize}

\subsubsection*{Bulk geometry}
The new inhomogeneous black brane solutions that we find have peculiar features, including
\begin{itemize}

\item The inhomogeneities are localized near the horizon, and die off asymptotically following a power law decay.

\item The phenomena of vorticity, frame dragging and the magneto-electric effect similar to one produced by a near horizon 
topological insulator are observed. 

\item  The inhomogeneous black brane has a neck and a bulge. In the curvature at the horizon, the maximum is at the bulge.
In the limit of small temperatures, the neck shrinks to zero size. 

\item The proper length of the horizon grows when temperature is decreasing, and diverges as $1/T^{0.1}$ in the limit $T\rightarrow 0$. The proper length in the stripe direction increases from the boundary to the horizon, which can be thought of as a manifestation of an ``Archimedes effect''.

\end{itemize}

In \S\ref{equations}, we define our model and set up our numerical approach, describing our ansatz, boundary conditions and solving procedure. Then, in \S\ref{sec_solutions}, we report on interesting geometrical features of the bulk solutions. \S\ref{sec_thermo_finite} studies the solutions at fixed length from the point of view of the boundary theory. There, we make the comparison to the homogeneous solution and find a second order transition, in addition to describing the observables in the theory. In \S\ref{sec_thermo_infinite}, we relax the fixed length condition and find the striped solution that dominates the thermodynamics for the infinite system. Appendix \ref{asympt_charges} provides details about computing the observables of the inhomogeneous solutions while appendix \ref{appendixB} gives more details on the numerics, including checks of the solutions and validations of our numerical method.

{\it Note added:} As this manuscript was being completed, \cite{Donos:2013wia} and \cite{Withers:2013} appeared, which use a different method and have some overlap with this work.

%====================================================================%
\section{Numerical set-up: Einstein-Maxwell-axion model}
\label{equations}

In \cite{Donos:2011bh}, perturbative instabilities of the Reissner-Nordstr\"{o}m-AdS (RN for short) black brane were found within the Einstein-Maxwell-axion model. In \cite{Stripes1} and here, we construct the full non-linear branch of stationary solutions following this zero mode. 

\subsection{The model and ansatz}

The Lagrangian describing our coupled system can be written as \cite{Donos:2011bh} \be\mathcal{L}= \half (R+12) - \half\d^\mu\psi \d_\mu\psi-\half m^2 \psi^2-\frac{1}{4} F^{\mu\nu}F_{\mu\nu} -\frac{1}{\sqrt{-g}}\frac{c_1}{16 \sqrt{3}} \,\psi \,\epsilon^{\mu\nu\rho \sigma}F_{\mu\nu}F_{\rho \sigma},\label{Lgrn}\ee where $R$ is the Ricci scalar,  $F_{\mu\nu}$ is the Faraday tensor, $\psi$ is a pseudo-scalar field and $g$ is the determinant of the metric. We use units in which the AdS radius $l^2=1/2$, Newton's constant $8 \pi G_N= 1$ and $c=\hbar=1$, and choose $m^2 =-4$. The constant $c_1$ controls the strength of the axion coupling. 

For this choice of scalar field mass, instabilities exist for all choices of $c_1$. For $c_1=0$, the instability is  towards a black hole with neutral scalar hair. For $c_1>0$, inhomogeneous instabilities along one field theory direction exist for a range of wavenumbers $k$. The critical temperature at which each mode becomes unstable depends on the wavenumber: $T_c(k)$. For a given $c_1$, there is a maximum critical temperature, above which there are no unstable modes. As one increases $c_1$, the critical temperature of a given mode $k$ increases, such that for a fixed temperature a larger range of wavenumbers will be unstable. See appendix \ref{lin_an} for more details on the perturbative analysis.

One may consider generalizations of this action, including higher order couplings between the scalar field and the gauge field. In particular, as discussed in \cite{Donos:2011bh}, generalizing the Maxwell term as $-\frac{\tau(\psi)}{4} F^{\mu\nu}F_{\mu\nu}$, where $\tau(\psi)$ is a function of the scalar field, results in a model that can be uplifted to a $D=11$ supergravity solution (for particular choices of $c_1$, $m$, and the parameters in $\tau(\psi)$). In this study, we wish to study the formation of holographic stripes phenomenologically. The existence of the axion-coupling term ($c_1\neq0$) is a sufficient condition for the inhomogeneous solutions and so we set $\tau(\psi)=1$ here.

We are looking for stationary black hole solutions that can be described by an ansatz of the form 
\be\label{metric}ds^2=-2r^2f(r)e^{2A(r,x)}dt^2+e^{2B(r,x)}\left(\frac{dr^2}{2r^2f(r)}+2r^2dx^2\right)+2r^2e^{2C(r,x)}(dy-W(r,x)dt)^2,\nn\ee\be\psi=\psi(r,x),\;\;\; A=A_t(r,x)dt+A_y(r,x)dy,\label{ansatz}\ee 
where $r$ is the radial direction in AdS and $x$ is the field theory direction along which inhomogeneities form. We term the scalar field and gauge fields collectively as the matter fields. $f(r)$ is a given function whose zero defines the black brane horizon. We take $f(r)$ to be that of the RN solution,\be f(r) = 1 - \left( 1 + \frac{\mu^2}{4r_0^2} \right) \left( \frac{r_0}{r} \right)^3 +\frac{\mu^2}{4r_0^2}\left( \frac{r_0}{r} \right)^4,\label{frn}\ee so that the horizon is located at $r=r_0$. The homogeneous solution is the RN black brane, given by \be A=B=C=W=\psi=A_y=0,\;\;\; A_t(r)=\mu(1-r_0/r),\ee
where $\mu$ is the chemical potential.  Above the maximum critical temperature, this is the only solution to the system.

To find the non-linear inhomogeneous solutions, we numerically solve the equations of motion derived from the ansatz (\ref{ansatz}). The Einstein equation results in four second order elliptic equations, formed from combinations of $G^t_t-T^t_t=0$, $G^t_y-T^t_y=0$, $G^y_y-T^y_y=0$, and $G^r_r+G^x_x-(T^r_r+T^x_x)=0$, and two hyperbolic constraint equations, $G^r_x-T^r_x=0$ and $G^r_r-G^x_x-(T^r_r-T^x_x)=0$, for the metric functions. The gauge field equations and scalar field equation give second order elliptic equations for the matter fields. For completeness, the full equations are given in appendix \ref{eom}. Our strategy will be to solve these seven elliptic equations subject to boundary conditions that ensure that the constraint equations will be satisfied on a solution. Below, we describe the constraint system and our boundary conditions. For more details about the numerical approach, we refer to appendix \ref{appendixB}.

%====================================================================%
\subsection{The constraints}

The two equations $G^r_x-T^r_x=0$ and $G^r_r-G^x_x-(T^r_r-T^x_x)=0$, which we do not explicitly solve, are the constraint equations. Using the Bianchi identities \cite{Wiseman:2002zc}, we see that the constraints satisfy \be \dx\left(\rootg (G^r_x-T^r_x)\right)+2r^2 \sqrt{f} \dr\left(r^2 \sqrt{f} \rootg (G^r_r - G^x_x-(T^r_r-T^x_x))\right)=0, \ee\be 2r^2\sqrt{f}\dr\left(\rootg (G^r_x-T^r_x)\right)-\dx\left(r^2 \sqrt{f} \rootg (G^r_r - G^x_x-(T^r_r-T^x_x))\right)=0. \ee Defining $\hat r$ by $\d_{\hat r}=2r^2\sqrt f\d_r$ gives Cauchy-Riemann relations \be \dx\left(\rootg (G^r_x-T^r_x)\right)+\d_{\hat r}\left(r^2 \sqrt{f} \rootg (G^r_r - G^x_x-(T^r_r-T^x_x))\right)=0, \ee\be\d_{\hat r}\left(\rootg (G^r_x-T^r_x)\right)-\dx\left(r^2 \sqrt{f} \rootg (G^r_r - G^x_x-(T^r_r-T^x_x))\right)=0,\ee showing that the weighted constraints satisfy Laplace equations. Then, satisfying one constraint on the entire boundary and the other at one point on the boundary implies that they will both vanish on the entire domain. In practice we will take either zero data or Neumann boundary conditions at the boundaries in the $x$-direction. The unique solution to Laplace's equation with zero data on the horizon and the boundary at infinity and these conditions in the $x$-direction is zero. Therefore, as long as we fulfill one constraint at the horizon and the asymptotic boundary and the other at one point (on the horizon or boundary), the constraints will be satisfied if the elliptic equations are. Our boundary conditions will be such that $\rootg (G^r_x-T^r_x)=0$ at the horizon and conformal infinity and that $r^2 \sqrt{f} \rootg (G^r_r - G^x_x-(T^r_r-T^x_x))=0$ at one point on the horizon.

%====================================================================%
\subsection{Boundary conditions}
\label{sec_bc}
\begin{figure}[t!] 
\begin{center}
\setlength{\unitlength}{1mm}
\begin{picture}(100,75)(0,-5)
        \linethickness{1.5pt}
      \includegraphics[width=10cm]{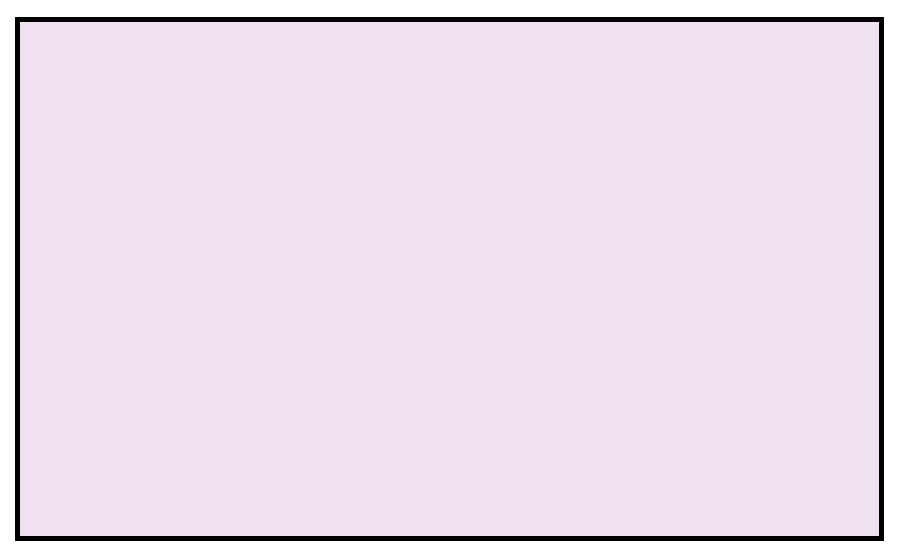}
      \multiput(-102, 1)(0, 4.9){12}{\line(1, 1){4}}

        \linethickness{1pt}
        \put(-98, 59){\vector(0, 1){10}}
        \put(-2, 2){\vector(1, 0){10}}

	\put(-117,33){\makebox(0,0){regularity,}}
	\put(-117,26){\makebox(0,0){$\sqrt{-g}G^r_x=0$}}
	
	\put(17,40){\makebox(0,0){$A,B,C,W\propto\frac{1}{r^3}$,}}
	\put(17,26){\makebox(0,0){$A_t-\mu,A_y\propto\frac{1}{r}$,}}
	\put(17,33){\makebox(0,0){$\psi\propto\frac{1}{r^2}$,}}
	\put(17,19){\makebox(0,0){$\sqrt{-g}G^r_x=0$}}
	
	\put(-50,64){\makebox(0,0){$\psi=\d_xA_y=\d_xg_{ty}=\d_xh=0$}}
	\put(-50,7){\makebox(0,0){$\d_x\psi=A_y=g_{ty}=\d_xh=0$}}
	
	\put(10,-2){\makebox(0,0){$r$}}
	\put(-2,-2){\makebox(0,0){$r=r_{cut}$}}
	\put(-97,-2){\makebox(0,0){$r=r_0$}}
	
	\put(-102,70){\makebox(0,0){$x$}}
	\put(-108,60){\makebox(0,0){$x=\frac{L}{4}$}}
	\put(-108,3){\makebox(0,0){$x=0$}}
\end{picture}
\caption{A summary of the boundary conditions on our domain. At the horizon, $r=r_0$, we impose regularity conditions. At the conformal boundary, $r\rightarrow\infty$, we have fall off conditions on the fields (imposed at large but finite $r=r_{cut}$) such that we do not source the inhomogeneity. In the $x$-direction, we use symmetries to reduce the domain to a quarter period $L/4$. Then, we impose either periodic or zero conditions on the fields, according to their behavior under the discrete symmetries discussed in the text. ($h$ collectively denotes the fields $\{g_{tt}, g_{xx}, g_{yy}, A_{t}\}$.) In addition to these, we explicitly satisfy the constraint equation $\sqrt{-g}G^r_x=0$ on the horizon and the conformal boundary.}
\label{fig_domain}
\end{center}
\end{figure}
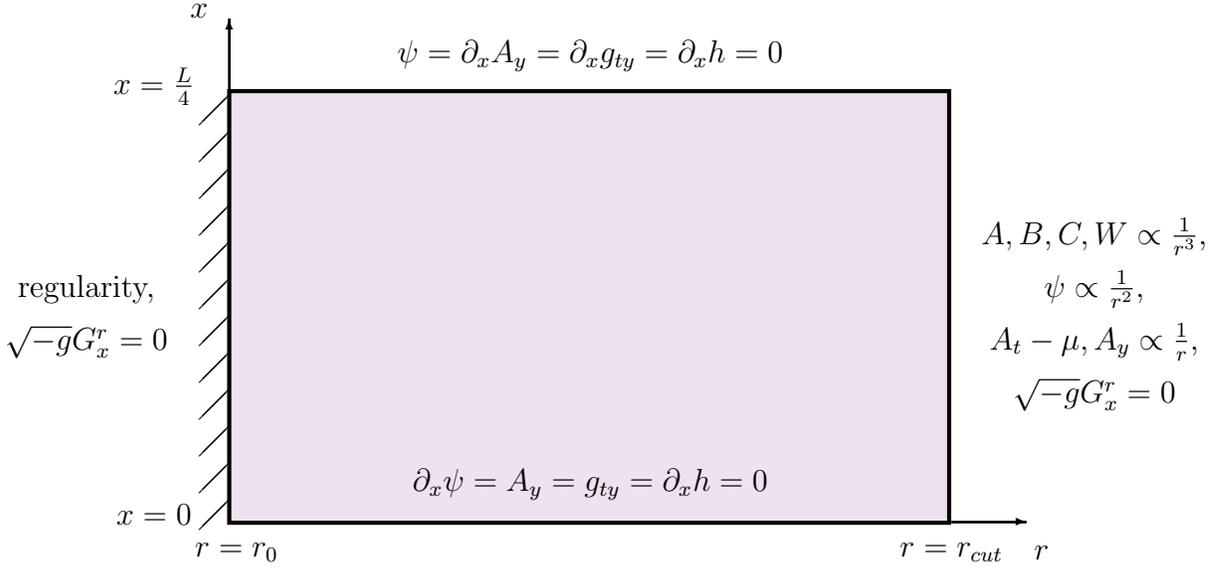

The elliptic equations to be solved are subject to physical boundary conditions. There are four boundaries of our domain (see Fig.~\ref{fig_domain}): the horizon, the conformal boundary, and the periodic boundaries in the $x$-direction, which are described next.

\subsubsection*{Staggered periodicity}
 To specify the boundary conditions in the $x$ direction we look at the form of the linearized perturbation which becomes unstable (see appendix \ref{lin_an}). To leading order in the perturbation parameter $\lambda$, they are of the form: \bea \psi(x) &\sim \lambda \cos (kx),\nn \\ A_{y}(x) &\sim \lambda \sin(kx),\nn \\ g_{ty}(x) &\sim \lambda \sin(kx),\eea where $k$ is the wavenumber of the unstable mode. To second order in the perturbation parameter, the functions $g_{tt}, g_{xx}, g_{yy}$ and $A_{t}$ (which we denote collectively as $h$) are turned on, with the schematic behavior \be h(x) \sim \lambda^2 (\cos(2kx) + C), \ee where $C$ are independent of $x$.
 
All these functions are periodic with period $L=\frac{2\pi}{k}$. However, they are not the most general periodic functions with period $L$. For numerical stability it is worthwhile to specify their properties further and encode those properties in the boundary conditions we impose on the full solution. We concentrate on the behavior of the perturbation with respect to two independent $Z_2$ reflection symmetries.
 
The first $Z_2$ symmetry is that of  $x \rightarrow {-x}$, $y \rightarrow -y$, which is a rotation in the $x,y$ plane. This is a symmetry of the action and of the linearized perturbation (keeping in mind that $A_y$ and $g_{ty}$ change sign under reflection of the $y$ coordinate). We conclude therefore that this is a symmetry of the full solution.

Similarly, the $Z_2$ operation $x \rightarrow \frac{L}{2} -x$, $y\rightarrow -y$ is a symmetry of the action, which is also a symmetry of the linearized system when accompanied by $\lambda \rightarrow -\lambda$. In other words the functions $\psi, A_y, g_{ty}$ are restricted to be odd with respect to this $Z_2$ operation, while the rest of the functions, which we collectively denoted as $h$, are even.

The two symmetries defined here restrict the form of the functions that can appear in the perturbative expansions for each of the functions above. For example, it is easy to see that the function $\psi(x)$ gets corrected only in odd powers of $\lambda$ and the most general form of the harmonic that can appear in the perturbative expansion is $\cos(nkx)$, for $n$ odd. Similar comments apply to the other functions above.

We restrict ourselves to those harmonics which may appear in the full solution. The most efficient way to do so is to work with a quarter of the full period $L$ (reconstructing the full periodic solution using the known behavior of each function with respect to the two $Z_2$ operations defined above). The specific properties of each function appearing in our solutions are imposed by demanding the following boundary conditions: \bea \partial_x\psi(x=0) &= &0,~~~~~~~~\psi(x=\frac{L}{4})=0, \nn \\ A_y(x=0) &=& 0,~~~~~~~~ \partial_x A_y(x=\frac{L}{4}) = 0,\nn \\ g_{ty}(x=0) &=& 0,~~~~ ~~~~\partial_x g_{ty}(x=\frac{L}{4}) = 0,\nn \\ \partial_x h(x=0) &=& 0,~~~~ ~~~~\partial_x h(x=\frac{L}{4}) = 0. \label{per_conds}\eea

\subsubsection*{At the horizon}
In our coordinates (\ref{metric})  the horizon is at fixed $r=r_0$. For numerical convenience we introduce another radial coordinate $\rho=\sqrt{r^2-r_0^2}$, such that the horizon is at $\rho=0$.\footnote{In the rest of the paper, we use $r$ and $\rho$ interchangeably as our radial coordinate. We use the coordinate $\rho$ in the numerics.} Expanding the equations of motion around $\rho=0$ yields a set of Neumann regularity conditions, \be\drho A=\drho C=\drho W = \drho \psi = \drho A_t = \drho A_y = 0, \ee and two conditions in the inhomogeneous direction along the horizon, \be\dx W = \dx(A_t+WA_y)=0. \label{horconds} \ee Thus, both $W$ and the combination $A_t+WA_y$ are constant along the horizon. The boundary conditions in the $x$ direction (\ref{per_conds}) imply that $W=0$. Then, the second condition together with regularity of the vector field $A$ on the Euclidean section give that $A_t=0$ on the horizon.

The regularity conditions give eight conditions for the six functions $A,C,W,\psi,A_t$ and $A_y$. In principle, we would choose any six of these to impose at the horizon. If we find a non-singular solution to the equations, then the other two conditions should also be satisfied. In practice, some of these conditions work better than others for finding the numerical solution. We find that using Neumann conditions for $A,C,\psi$, and $A_y$ and Dirichlet conditions for $W$ and $A_t$ results in a more stable relaxation.\footnote{Using Neumann conditions at the horizon for $W$ and $A_t$ results in values at the horizon that converge to zero with step-size, consistent with the above analysis.}

The conditions for $B$ are determined using the constraint equations. Expanding the weighted constraints at the horizon, we find \bea \rootg (G^r_x-T^r_x)&\propto&\dx(A-B)+O(\rho),\\ r^2 \sqrt{f} \rootg (G^r_r - G^x_x-(T^r_r-T^x_x))&\propto& \drho B+ O(\rho). \eea The first condition gives constant surface gravity (or temperature) along the horizon. As discussed above, we will impose one constraint at the horizon and the boundary, and the other at one point. In practice, we will satisfy $r^2 \sqrt{f} \rootg (G^r_r - G^x_x-(T^r_r-T^x_x))$ at $(\rho,x)=(0,0)$, updating the value of $B$ at this point using the Neumann condition $\drho B=0$. This will set the difference $(B-A)|_{(\rho,x)=(0,0)}\equiv d_0$, which we will then use to update $B$ using a Dirichlet condition along the rest of the horizon, satisfying $\rootg (G^r_x-T^r_x)=0$.

\subsubsection*{At the conformal boundary}

In our coordinates, the boundary is at $r=\infty$. Since we are looking for spontaneous breaking of homogeneities, our boundary conditions will be such that the field theory sources are homogeneous. This implies that the non-normalizable modes of the bulk fields are homogeneous. The inhomogeneity of the striped solutions will be imprinted on the normalizable modes of the fields, or the coefficient of the next-to-leading fall-off term in the asymptotic expansions.

The form of our metric ansatz is such that the metric functions $A,B,C$ and $W$ represent the normalizable modes of the metric. Imposing that the geometry is asymptotically AdS with Minkowski space on the boundary implies that these four metric perturbations must vanish as $r\rightarrow\infty$. By expanding the equations of motion near the boundary, one can show that $A,B,C$ and $W$ fall off as $1/r^3$. In practice, we place the outer boundary of our domain at large but finite $r_{cut}$ and impose the fall-off conditions there.  

As in the RN solution, we source the field theory charge density with a homogeneous chemical potential, corresponding to a Dirichlet condition for the gauge field $A_t$ at the boundary. In the inhomogeneous solutions, we expect the spontaneous generation of a modulated field theory current $j_y(x)$, dual to the normalizable mode of $A_y$. Solving the equations near the boundary with these conditions reveals the expansions $A_t=\mu+O(1/r)$ and $A_y=O(1/r)$, which we impose numerically at $r_{cut}$.

The scalar field equation of motion gives the asymptotic solution \be \psi=\frac{\psi^{(1)}}{r^{\lambda_-}}+\frac{\psi^{(2)}}{r^{\lambda_+}}+\dots,\ee where \be \lambda_\pm=\half\left( 3\pm \sqrt{9+4(lm)^2} \right).\ee For the range of scalar field masses $-9/2\le m^2\le -5/2$, both modes are normalizable, and fixing one mode gives a source for the other. In our study we will choose $m^2=-4$, giving $\lambda_-=1$, $\lambda_+=2$. Since we are looking for spontaneous symmetry breaking, in this case we must choose either $\psi^{(1)}=0$ or $\psi^{(2)}=0$. We choose the former, so that $\psi$ falls off as $1/r^2$.

Now, consider the weighted constraint $\rootg G^r_x$. As discussed above, in order to solve the constraint system, we require this to disappear at the conformal boundary. Near the boundary, $\rootg\propto r^2+\dots$, so for $\rootg G^r_x$ to disappear we must have $G^r_x=O(1/r^3)$. Expanding the equations near the boundary we have \be G^r_x-T^r_x\propto\frac{3\dx A^{(3)}(x)+2\dx B^{(3)}(x)+3\dx C^{(3)}(x)}{r^2}+O\left(\frac{1}{r^3}\right),\label{asymp_con}\ee where $X=X^{(3)}(x)/r^3+\dots$ for $X=\{A,B,C\}$. Therefore, in addition to the boundary conditions mentioned above, for $\rootg G^r_x=0$ to be satisfied at $r=\infty$, it appears that we should have that $3A^{(3)}(x)+2B^{(3)}(x)+3C^{(3)}(x)=const$. The means to impose this addition condition comes from the fact that our metric (\ref{metric}) has an unfixed residual gauge freedom \cite{Aharony:2005bm}, allowing one to transform to new $\tilde{r}=\tilde{r}(r,x), \tilde{x}=\tilde{x}(r,x)$ coordinates which are harmonic functions of $r$ and $x$. Performing such a transformation generates an additional function in (\ref{asymp_con}), which can then be chosen to ensure that the constraint is satisfied (in appendix \ref{appendixB} we describe how). This condition implies the conservation of the boundary energy momentum tensor, see appendix \ref{asympt_charges}.

%========================================================
\subsection{Parameters and algorithm}

The physical data specifying each solution is the chemical potential $\mu$, the temperature $T$, and the periodicity $L$.\footnote{Fixing $\mu$, $T$ and $L$ gives the system in the grand canonical ensemble. Once the phase space has been mapped in one ensemble other ensembles can be considered via appropriate reinterpretation of the numerical data. See \S\ref{sec_thermo_finite} for a description of this process.} Since the boundary theory is conformal, it will only depend on dimensionless ratios of these parameters. This manifests itself in the following scaling symmetry of the equations: 
\bea r\rightarrow \lambda r, \;\; (t,x,y)\rightarrow\frac{1}{\lambda}(t,x,y), \;\; A_\mu\rightarrow \lambda A_\mu. 
\label{scaling_symmetry}
\eea
 We use this to select $\mu=1$. Then, our results are functions of the dimensionless temperature $T/\mu$ and the dimensionless periodicity $L\mu$.

The temperature is controlled by the coordinate location of the horizon. For a given $r_0$, the temperature of the RN phase is $T_0=(1/8\pi r_0)(12r_0^2-1)$ while the temperature of the inhomogeneous solution is $T=e^{-d_0}T_0$. Recall that $(B-A)|_{r_0}=d_0$ is dynamically generated by satisfying the constraints at the horizon. From our numerical solutions, we find that $d_0$ monotonically increases as we lower the temperature, so that $T_0$ gives a reliable parametrization of the physical temperature $T$. In practice, we generate solutions by choosing values of $T_0$ below the critical temperature $T_c(k)$.

We solve the equations by finite-difference approximation (FDA) techniques. We use second order FDA on the equations (\ref{eqA}) - (\ref{eqAy}) before using a point-wise Gauss-Seidel relaxation method on the resulting algebraic equations. 
For the results in this paper, for $c_1=4.5$, a cutoff of $\rho_{cut}=\{6,8\}$ was used while for $c_1=5.5$ and $c_1=8$, for which the modulations were larger, a cutoffs of $\rho_{cut}=10$ and $\rho_{cut}=12$ correspondingly were used. Grid spacings used for the FDA scheme were in the range $d\rho, dx=0.04-0.005$. Neumann boundary conditions are differenced to second order using one-sided FDA stencils in order to update the boundary values at each step. At the asymptotic boundary $\rho_{cut}$ we impose the boundary conditions by second order differencing a differential equation based on the fall-off (for example, $\d_rA=-3A/r$) to obtain an update rule for the boundary value. As a result we find quadratic convergence as a function of grid-spacing for our method, see appendix \ref{appx_conv}.

%=====================================================
\section{The solutions}
\label{sec_solutions}
The system of equations (\ref{eqA}-\ref{eqAy}) is solved subject to
boundary condition described in the previous sections. The details of
our numerical algorithm are found in appendix \ref{appendixB}. Here we
focus on the properties of the solutions and their geometry.  

Unless otherwise specified the following plots were obtained using the axion coupling of $c_1=4.5$.
In this section, we consider solutions for which the periodicity is determined by the dominant critical wavenumber $k_c$; for $c_1=4.5$, this gives $L\mu/4\simeq2.08$, see Table \ref{c1_Tc}.
We found that the geometry and most of the other features are qualitatively similar for the 
couplings $c_1=5.5$ and $c_1=8$. 
A convenient way to parametrize our inhomogeneous solutions is by
the dimensionless temperature  $T/T_c$, relative to the critical
temperature $T_c$, below which the translation invariance along $x$ is
broken. For $c_1=4.5$, our method allows us find solutions in the
range $0.003 \lesssim T/T_c \lesssim 0.9$.   

%======================
\subsection{Metric and fields}
\label{sec_metric}

\begin{figure}[t!] 
\center
    \includegraphics[width=14cm]{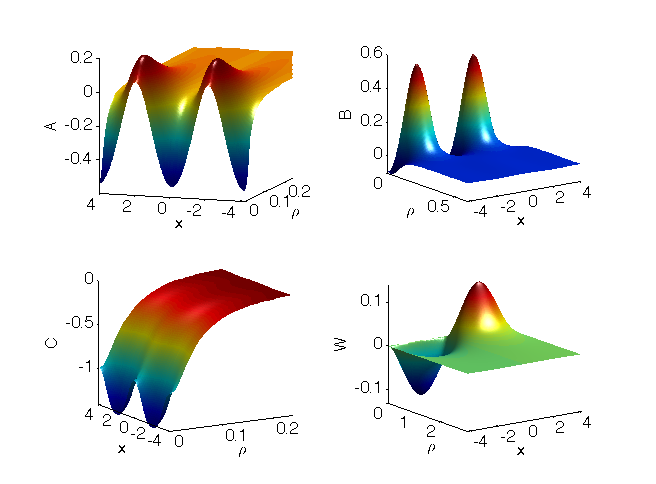}
    \caption[]{Metric functions for $T/T_c\simeq 0.11$. Note the
      metric functions $A, B$ and $C$ have half the period of $W$. The
    variation is maximal near the horizon, located at $\rho=0$, and it
  decays  as the conformal boundary is approached, when $\rho
  \rightarrow \infty$.}
\label{fig_ds_3D}
\end{figure}

\begin{figure}[t!] 
\center
    \includegraphics[width=15cm]{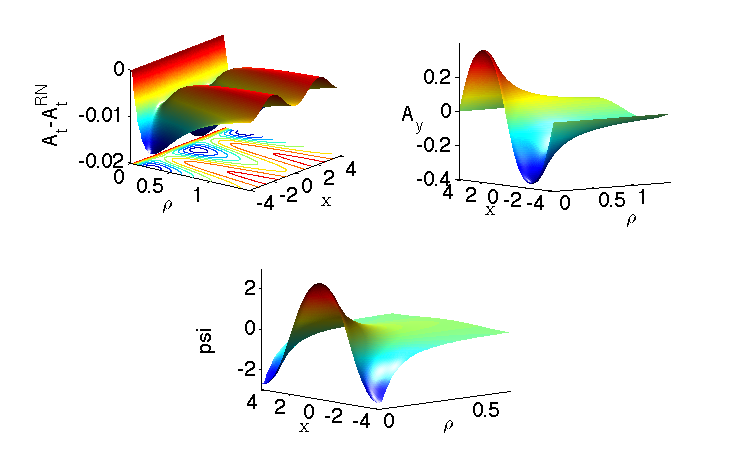}
    \caption[]{$A_t$ relative to the corresponding RN solutions, $A_y$
      and $\psi$ for $T/T_c\simeq 0.11$. The period of $A_t$ is
      twice that of $\psi$ and $A_y$.  The $x$-dependence dies off
      gradually as the conformal boundary is approached, at $\rho
      \rightarrow \infty$.  }
\label{fig_matter_3D}
\end{figure}

For subcritical temperatures, as we descend into inhomogeneous regime, the
metric and the matter fields start developing increasing variation in $x$.
Fig.~\ref{fig_ds_3D} displays the metric functions, 
and Fig.~\ref{fig_matter_3D} shows the non vanishing components of the vector
potential field and of the scalar field for $T/T_c\simeq
0.11$ over a full period in the $x$ direction. The variation of all
fields is maximal near the horizon of the black hole at $\rho=\sqrt{r^2-r_0^2}=0$, and
it gradually decreases toward the conformal boundary, $\rho
\rightarrow \infty$. 

\begin{figure}[t!] 
\center
    \includegraphics[width=12cm]{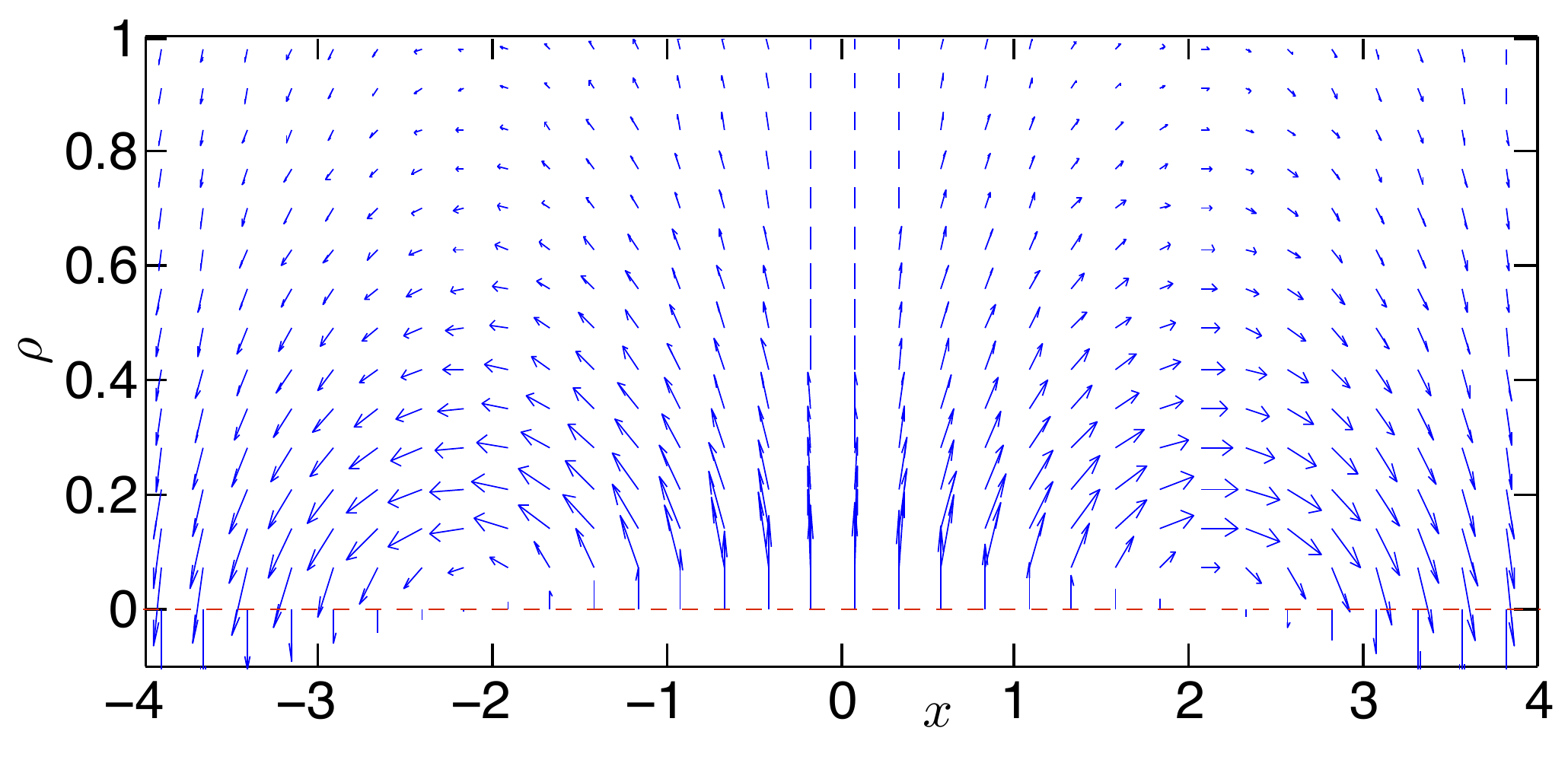}
    \caption[]{Magnetic field lines for solution with $T/T_c
      \simeq 0.07$. The pattern of vortices of alternating field
      directions form at the horizon (located at $\rho=0$).   }
\label{fig_magnetic_field}
\end{figure}

Many of the special features of the solutions we find may be explained via axion electrodynamics as seen in the effective description of
the electromagnetic response of a topological insulator. This effect is mediated by the interaction
term in our Lagrangian (\ref{Lgrn}). In the broken phase we have an axion gradient in the
near horizon geometry, which realizes a topological
insulator interface, see Fig.~\ref{fig_matter_3D}.
The characteristic patterning of the near horizon
magnetic field, $\mathbf{B} = \nabla \times \mathbf{A}$, shown in Fig.~\ref{fig_magnetic_field},  is reminiscent of
the magnetoelectric effect at such interfaces. The magnetic vortices are localized
near the black hole horizon and have alternating direction of magnetic
field lines.     

In curved space the magnetic field is accompanied by vorticity, which
is manifested by the function $W$. This causes frame dragging effects in
the $y$ direction. Test particles will be pushed along $y$ with speeds
$W(r,x)$, in particular the direction of the flow reverses every half
the period along $x$. The drag vanishes at the horizon and at
the location of the nodes of $W$ where $x= n\,L/2$, for integer $n$,
see Fig.~\ref{fig_ds_3D}.  In general,  the dragging effect remains bounded, and no ergoregion forms,
where the vector $\d_t$ becomes spacelike.

%==============================================
\subsection{The geometry}
\label{sec_geometry}
There are several ways to envisage the geometry of
our solutions, we discuss them in turn.  

\begin{figure}[t!] 
\center
    \includegraphics[width=9cm]{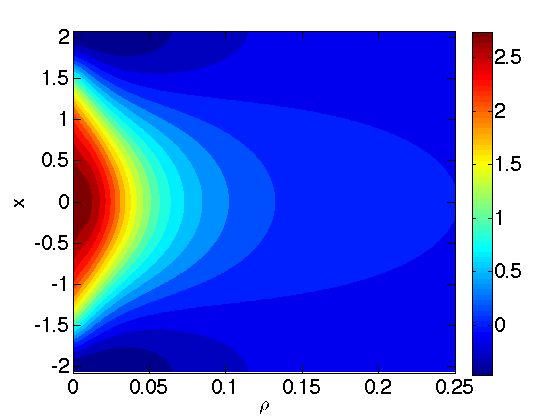}
    \caption[]{Ricci scalar relative to that of RN black hole, $R/R_{RN}-1$,
      $R_{RN}=-24$,  for $T/T_c\simeq 0.054$ over half the period.
      The scalar curvature is maximal along the horizon at $x=n\,
      L/2$ for integer $n$. }
\label{fig_Ricci}
\end{figure}

  The Ricci scalar of the RN solution is $R_{RN}=-24$, constant in $r$ and
  independent of the parameters of the
  black hole. This is no longer true for the inhomogeneous phases, where
  the Ricci scalar becomes position dependent. Fig.~\ref{fig_Ricci} 
  illustrates the spatial variation of the Ricci
  scalar, relative to the $R_{RN}$ for $T/T_c\simeq 0.054$. The
  maximal curvature is always along the horizon at  $x=n\, L/2$ for
  integer $n$. It grows when the temperature decreases and approaches
  the finite value of $R \simeq -94$ in the small temperature limit.  

\begin{figure}[t!] 
\center
    \includegraphics[width=10cm]{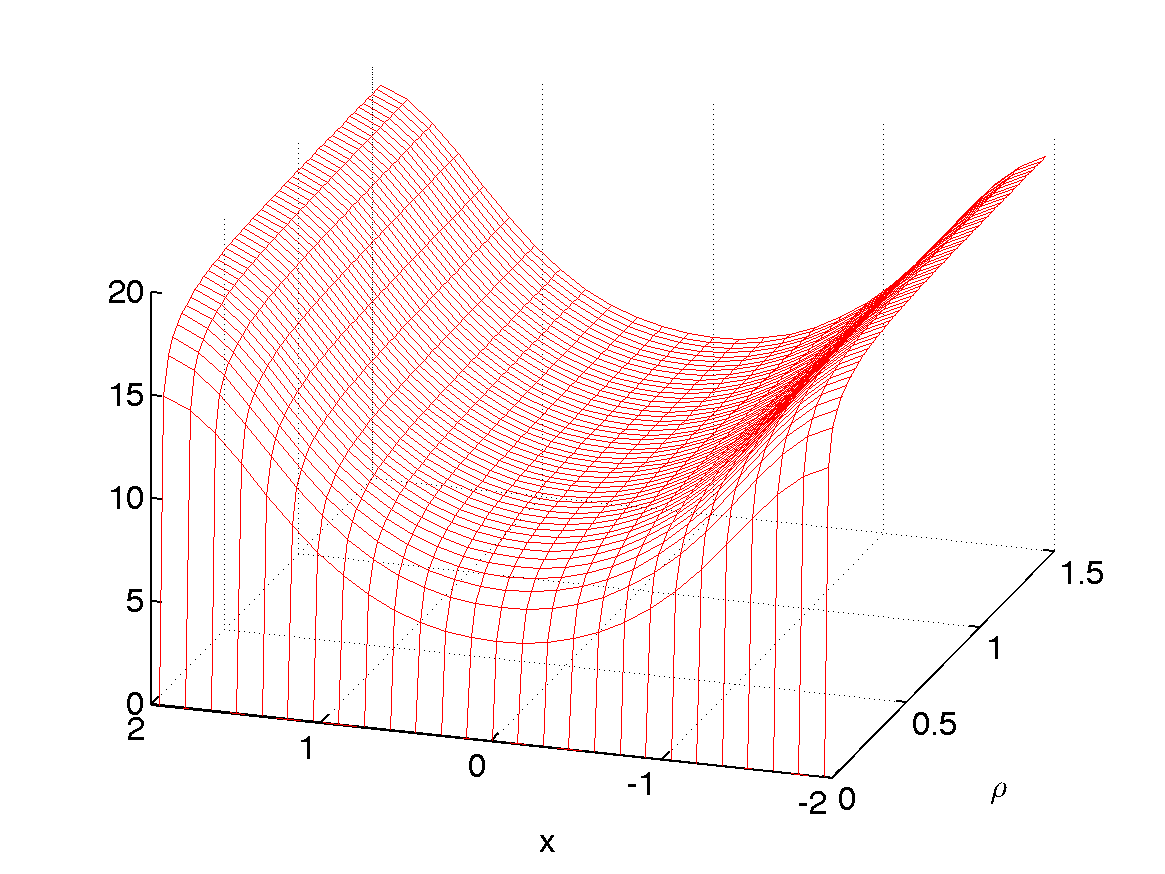}
    \caption[]{The embedding diagram of constant $x$ spatial slices,
      as a function of $x$ at given $y$ for $T/T_c\simeq0.035$. The
      geometry of $\rho=const$ slices is maximally curved at $x=n\,
      L/2$ for integer $n$.  }
\label{fig_embedding}
\end{figure}

Embedding in a given background
space is a convenient way to illustrate curved geometry. We consider
the embedding of 2-dimensional spatial slices of constant $x$ of the
full geometry (\ref{metric}) 
\be
\label{ds_2}
ds^2_2=\frac{e^{2\,B(r,x)}}{2\, r^2\, f(r)} dr^2 +2\, r^2\, e^{2\,
  C(r,x)}dy^2
\ee
as a surface in 3-dimensional AdS space
\be
ds_3^2=2\, \tlr^2\,dz^2+\frac{d\tlr^2}{2\, \tlr^2} + 2\,\tlr^2 dy^2. 
\ee
We are looking for a hypersurface parametrized by  $z=z(\tlr)$. Then 
the metric on such a hypersurface reads
\be
\label{ds_emb}
ds_2^2=\[1+2\,\tlr^2\(\frac{dz}{d\tlr}\)^2\]\frac{d\tlr^2}{2\, \tlr^2}+ 2\,\tlr^2 dy^2.
\ee
Comparing (\ref{ds_emb}) and (\ref{ds_2}) we obtain set of the relations
\bea
\label{embed_coords}
 \tlr&=&r\, e^{C},  \nonumber \\
  \[\frac{1}{2\, \tlr^2}+2\,\tlr^2\(\frac{dz}{d\tlr}\)^2\]\(\frac{d\tlr}{dr}\)^2&=&\frac{e^{2\,B(r,x)}}{2\, r^2\, f(r)} ,
\eea
resulting in the embedding equation
\be
\label{embedding}
\frac{dz}{dr} = \frac{1}{2 \, r^2} \sqrt{f(r)^{-1}\,e^{2\, B(r,x) -2\, C(r,x)
  }-\(1+r\, \d_r C(r,x)\)^2}.
\ee
We integrate this equation for a given $x$, and in Fig.~\ref{fig_embedding} 
show the embedding at constant $y$. The maximal
curvature along $\rho=const$ slices occurs at $x=n\,
      L/2$ for integer $n$, which is consistent with Fig.~\ref{fig_Ricci}.

\begin{figure}[t!] 
\center
    \includegraphics[width=8cm]{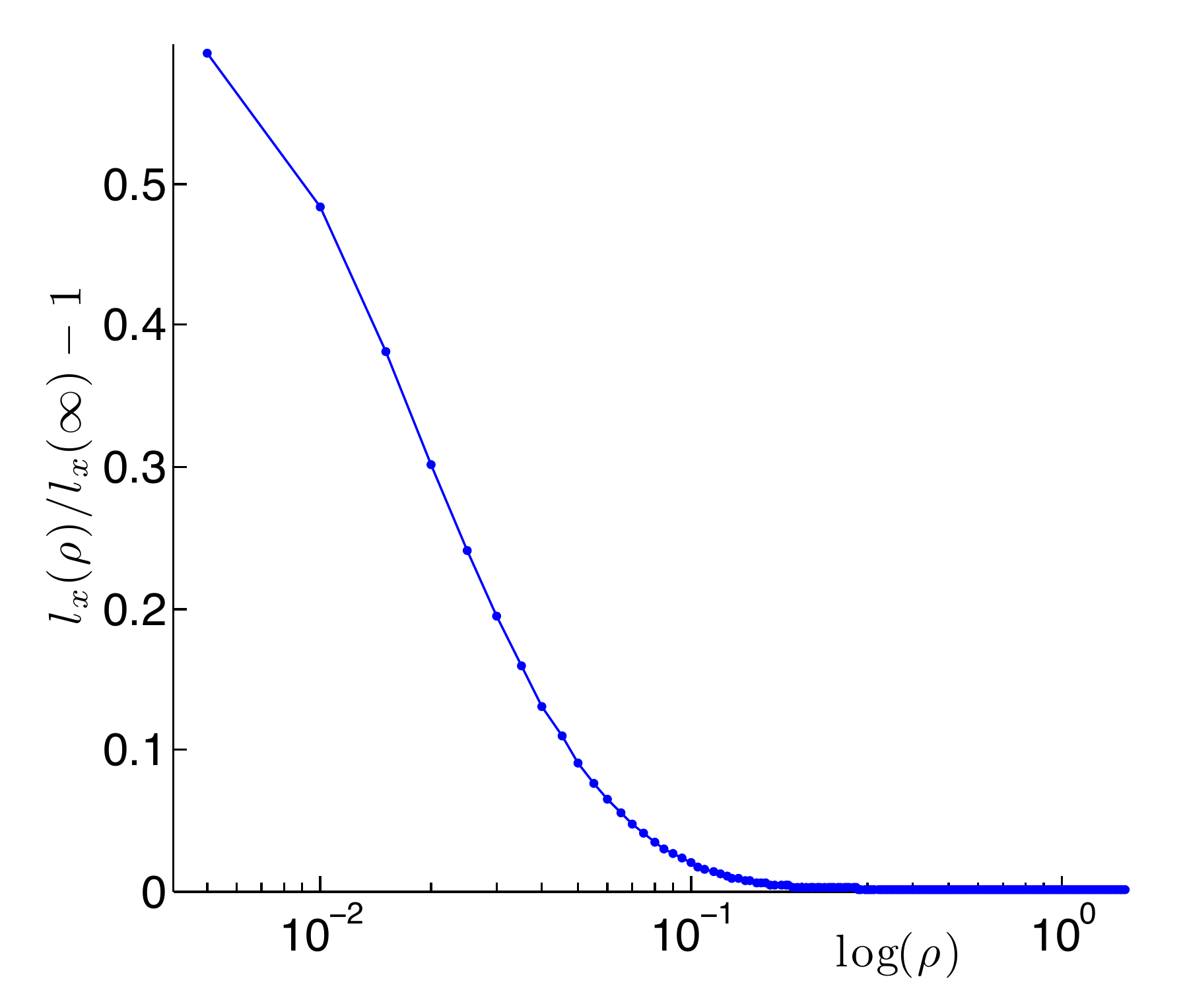}
    \caption[]{Radial dependence of the normalized proper length along
      $x$ for  $T/T_c\simeq 0.054$. While asymptotically the proper
      length coincides with the coordinate size of the strip, it grows
    as the horizon is approached. This is a manifestation of the ``Archimedes effect''.}
\label{fig_lx_r}
\end{figure}

The proper length of the stripe along $x$ relative to the background AdS spacetime at given
$r$ is  
\be
\label{lx}
l_x(r)/l_x(r=\infty) = \int_0^{L/4} e^B(r,x)\,dx.
\ee
Fig.~\ref{fig_lx_r} shows the dependence of the normalized proper length on the
radial distance from the horizon.
The proper length tends to the
coordinate length as $1/r^3$ asymptotically as $r \rightarrow \infty$,
but it exceeds that as the horizon is approached. Namely, the
inhomogeneous black brane ``pushes space'' around it
along $x$, in a manner resembling the ``Archimedes effect''.

\begin{figure}[t!] 
\center
   \includegraphics[width=9cm]{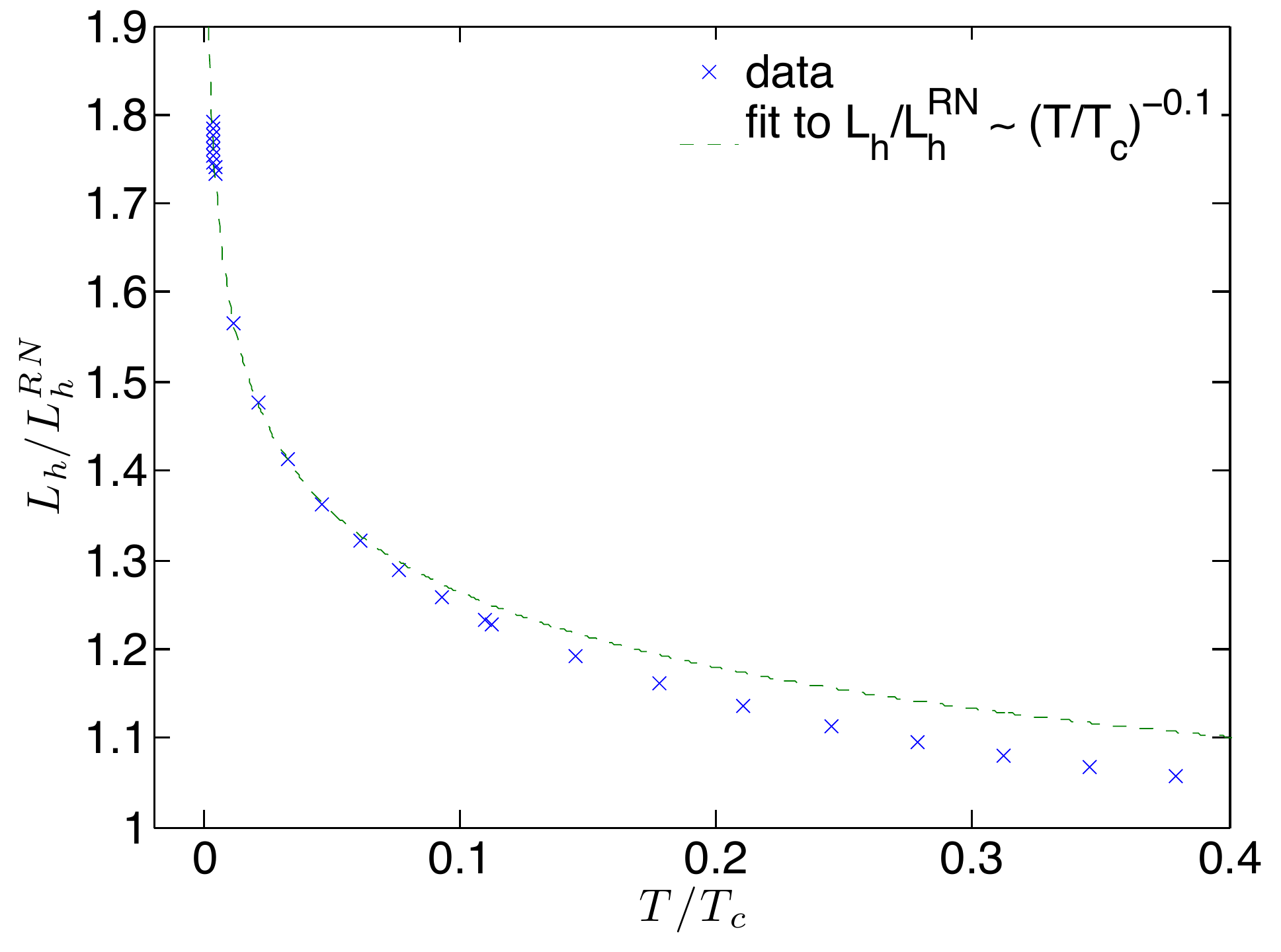}
    \caption[]{Temperature dependence of the proper length of the
      horizon along the stripe. Starting from as low as $L$ at high
      temperatures, the proper length grows monotonically and for small
      $T/T_c$ the growth is well approximated by the power-law
      dependence $ \sim (T/T_c)^{-0.1}$. }
\label{fig_lhor_T}
\end{figure}

The proper length of the horizon in $x$ direction is obtained calculating
(\ref{lx}) at $r_0$. Fig.~\ref{fig_lhor_T} demonstrates the dependence
of  this quantity on the temperature. For high temperatures the
length of the horizon resembles that of the homogeneous RN solution,
however, it grows when temperature
decreases.  We find that at small $T/T_c$
the proper length of the horizon diverges approximately as $(T/T_c)^{-0.1}$.

\begin{figure}
\center
    \includegraphics[width=10cm]{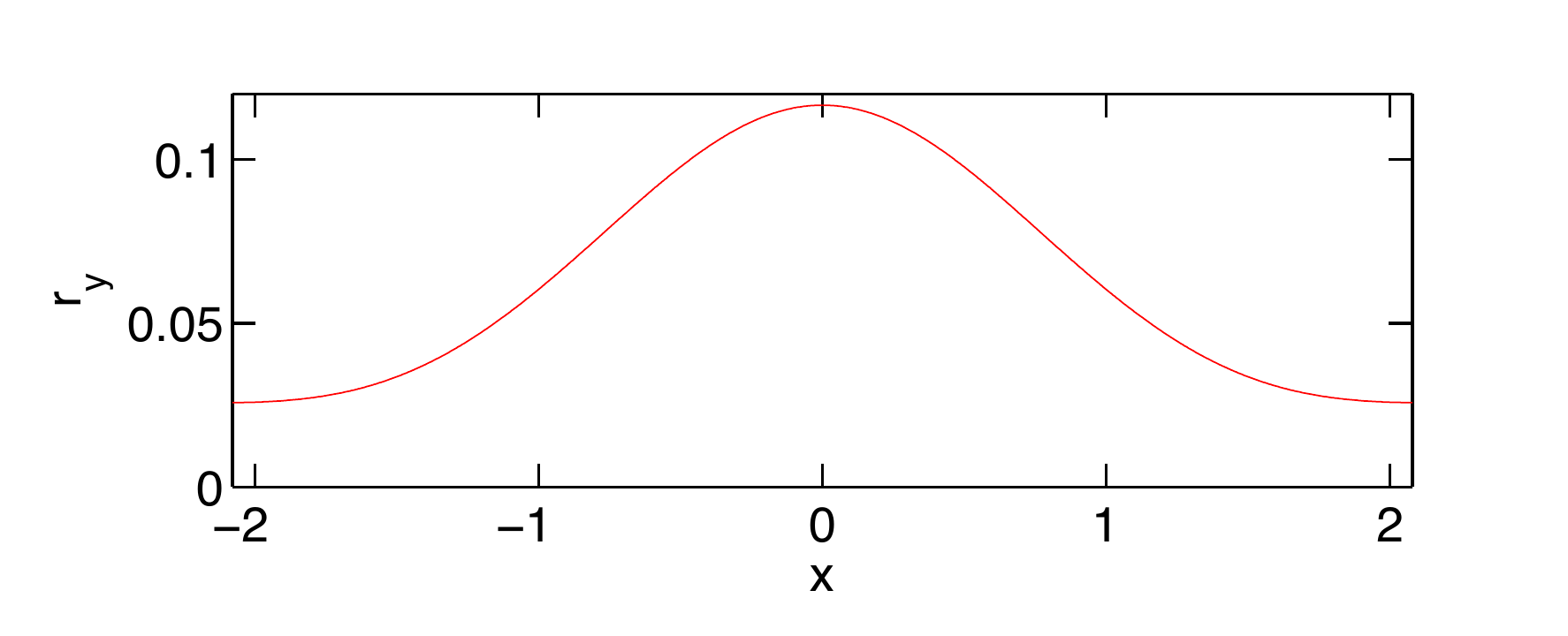}
    \caption[]{The the extent of the horizon in the transverse
      direction,  $r_y$,  as a functions of $x$ for
      $T/T_c\simeq0.054$ in $x \in [-L/2, L/2]$.  The characteristic pattern of
      alternating ``necks'' and ``bulges'' forms along $x$.}
\label{fig_ry}
\end{figure}

\begin{figure}
\center
    \includegraphics[width=9cm]{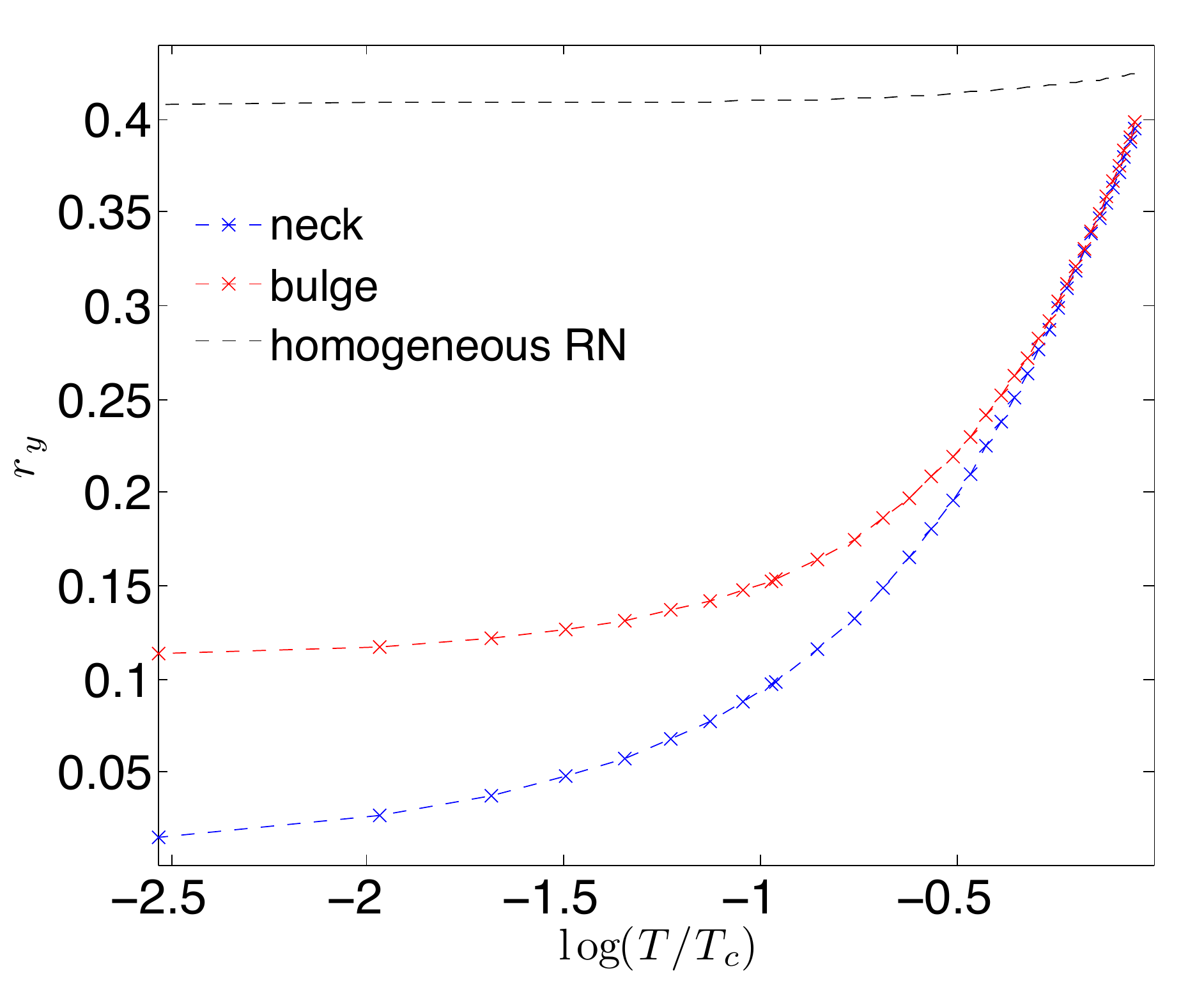}
    \caption[]{The dependence of the size of the neck and the bulge on
    temperature. }
\label{fig_ry_T}
\end{figure}

\begin{figure}
\center
   \includegraphics[width=9cm]{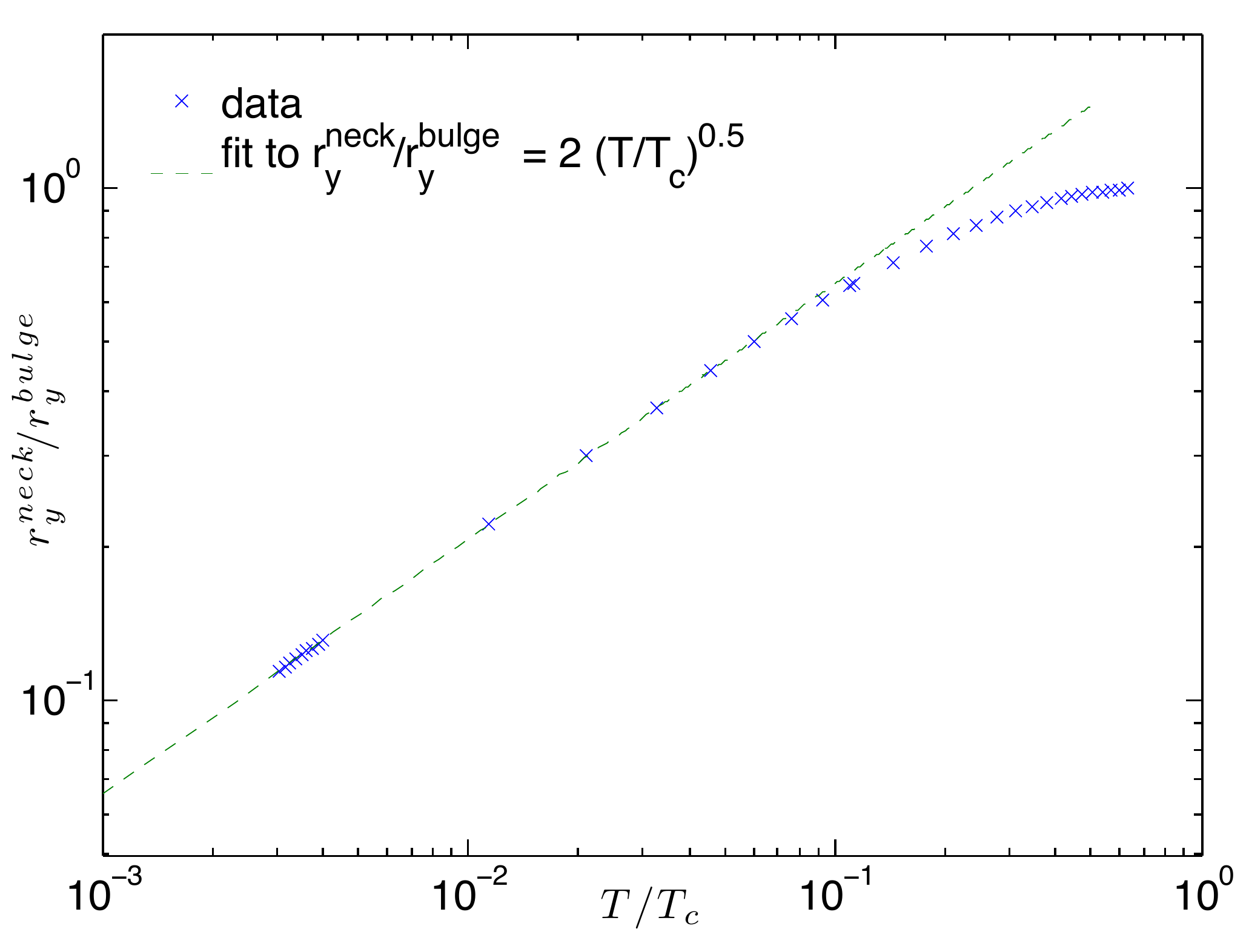}
                \caption{The ratio of the transverse extents of the
                  neck and the bulge shrinks as
    $r_y^{neck}/r_y^{bulge} \sim (T/T_c)^{1/2}$ at small temperatures, indicating a pinch-off of the horizon in the limit $T \rightarrow 0$.}
\label{fig_ry_min-max}
\end{figure}

The transverse extent of the horizon, per unit coordinate length $y$, is given by
\be
\label{ry}
r_y(x) = \sqrt{2}\, r_0\, e^{C(r_0,x)}.
\ee
Fig.~\ref{fig_ry} shows the variation of $r_y(x)$ along the horizon for
$T/T_c\simeq0.054$. Typically there is a ``bulge''  occurring  at $x=
n\, L/2$ and a ``neck'' at  $x= (2\,n+1)\, L/4$, for integer $n$.
Comparing this with 
Fig.~\ref{fig_Ricci} we note 
that Ricci scalar curvature is maximal at the bulge and not at the neck
as would happen, for instance, in the cylindrical geometry in black string case \cite{Sorkin:2006wp}.
Fig.~\ref{fig_ry_T} displays the dependence of the sizes of the neck and
bulge on $T/T_c$. Both sizes monotonically decrease with
temperature, however the rate at which the neck is shrinking exceeds
that of the bulge. This is demonstrated in Fig.~\ref{fig_ry_min-max}. In fact, we find that for $c_1=4.5$, $r_y^{neck}/r_y^{bulge} \sim
(T/T_c)^{1/2}$ near the lower end of the range of temperatures that we investigated. For other values of the axion coupling the scaling of 
the ratio is again power-law, with an exponent of the same order of magnitude, e.g. for $c_1=8$, the exponent is about $0.12$. This signals
a pinch-off of the horizon in the limit $T \rightarrow 0$. 

%===========================================================%
\section{Thermodynamics at finite length}
\label{sec_thermo_finite}
In this section we consider the thermodynamics and phase transitions in the system, assuming that the stripe length is kept fixed. For the finite system the length of the interval is part of the specification of the ensemble and is kept fixed.  In the next section we discuss the infinite system, for which the stripe width can adjust dynamically.
\subsection{The first law}

We demonstrated that below the critical temperature there exists a new branch of solutions which are spatially inhomogeneous. In the microcanonical ensemble the control variables of the field theory are the entropy $S$, the charge density $N$, and the length of the $x$-direction $L$, with corresponding conjugate variables temperature $T$, chemical potential $\mu$, and tension in the $x$-direction $\tau_x$\footnote{Explicit expressions for these quantities in terms of our ansatz are given in appendix \ref{asympt_charges}.}. The usual first law is augmented by a term corresponding to expansions and contractions in the $x$-direction and is given by
 \be dM=TdS+\mu dN +\tau_x dL. \label{first}\ee 
where $M$, $S$, and $N$ are quantities per unit length in the trivial $y$ direction, but are integrated over the stripe.

Our system has a scaling symmetry given by (\ref{scaling_symmetry}). In the field theory, this corresponds to a change of energy scale. Under this transformation, the thermodynamic quantities scale as \bea M\rightarrow\lambda^2 M, \;\;\; T\rightarrow\lambda T, \;\;\; S\rightarrow\lambda S, \;\;\; \mu\rightarrow\lambda \mu,\nn\\ N\rightarrow\lambda N, \;\;\;\;\; \tau_x\rightarrow\lambda^3 \tau_x, \;\;\;\;\; L\rightarrow \frac{1}{\lambda}L.\;\;\;\;\;\; \eea Using these in (\ref{first}) with $\lambda=1+\epsilon$, for $\epsilon$ small, yields\be 2M=TS+\mu N -\tau_x L,\label{ward}\ee the Smarr's-like expression that our solutions must satisfy and that can be used as a check of our numerics. For all of our solutions, we have verified that this identity is satisfied to one percent.

\subsection{Phase transitions}

The question of which solution dominates the thermodynamics depends on the ensemble considered. In the holographic context the choice of thermodynamic ensemble is expressed through the choice of boundary conditions. The corresponding thermodynamic potential is computed as the on-shell bulk action, appropriately renormalized and with boundary terms rendering the variational problem well-defined. We examine each ensemble in turn.

\subsubsection{The grand canonical ensemble}

In our numerical approach, the natural ensemble to consider is the grand canonical ensemble, fixing the temperature $T$, the chemical potential $\mu$, and the periodicity of the asymptotic $x$ direction as $L$. The corresponding thermodynamic potential is the grand free energy density \be \Omega(T,\mu,L)= M-TS-\mu N. \ee Different solutions of the bulk equations with the same values of $T,\mu,L$ correspond to different saddle point contributions to the partition function. The solution with smallest grand free energy $\Omega$ is the dominant configuration, determining the thermodynamics in the fixed  $T,\mu,L$ ensemble. In our case we have two solutions for each choice of $T,\mu,L$, one homogeneous and one striped. 
Exactly how one one saddle point comes to dominate over the other at temperatures below the critical temperature determines the order of the phase transition.

\begin{figure}
\center
    \includegraphics[width=15cm]{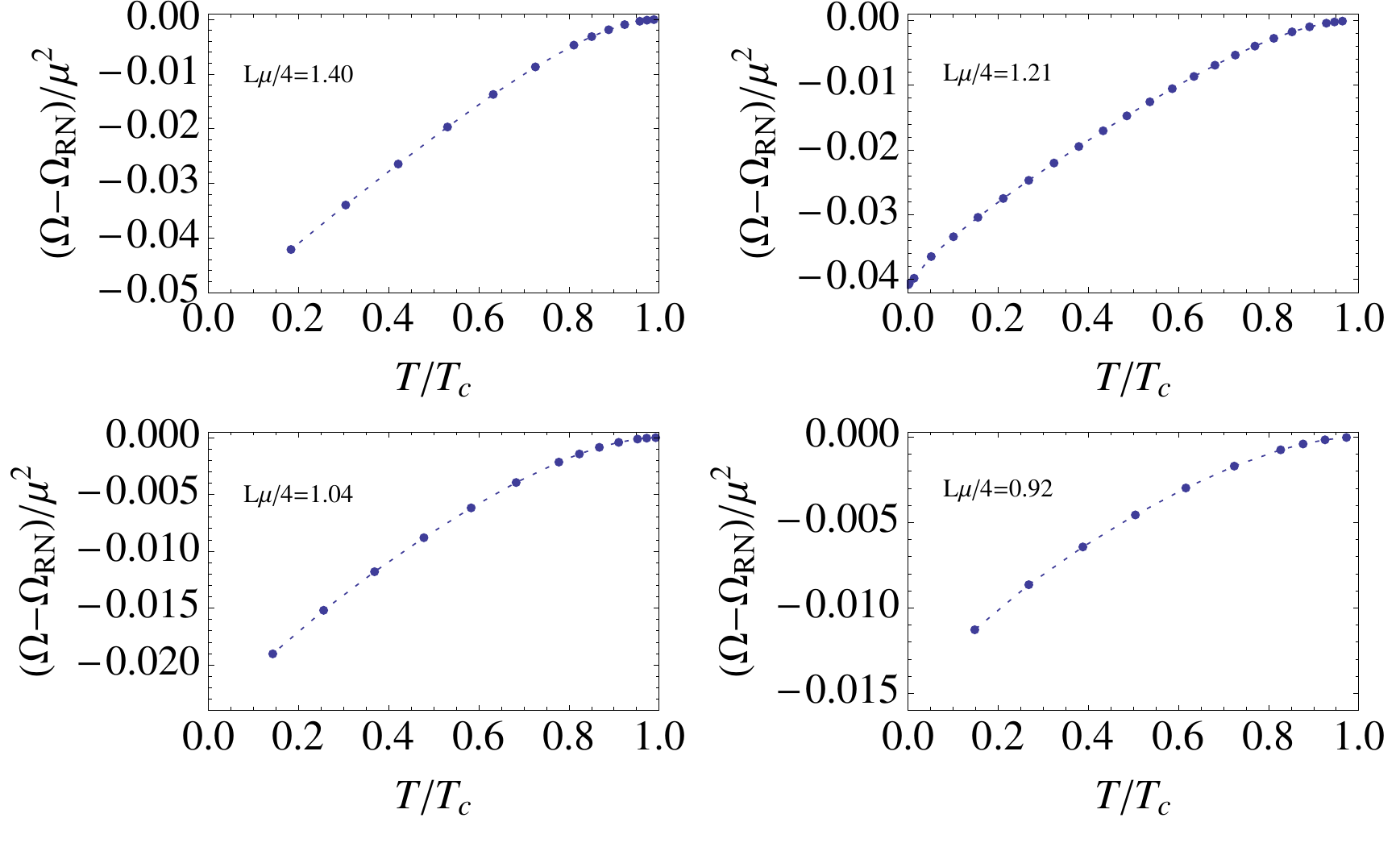}
    \caption[]{The grand free energy relative to the RN solution for several solutions of different fixed lengths at $c_1=8$.
In all cases shown we observe a second order phase transition. The critical exponents determined near the critical points in each case are consistent with the quadratic behavior $(\Omega-\Omega_{RN})/\mu^2 \propto (1-T/T_c)^2$.
}
\label{omegadiffs}
\end{figure}

\begin{figure}
\center
    \includegraphics[width=10cm]{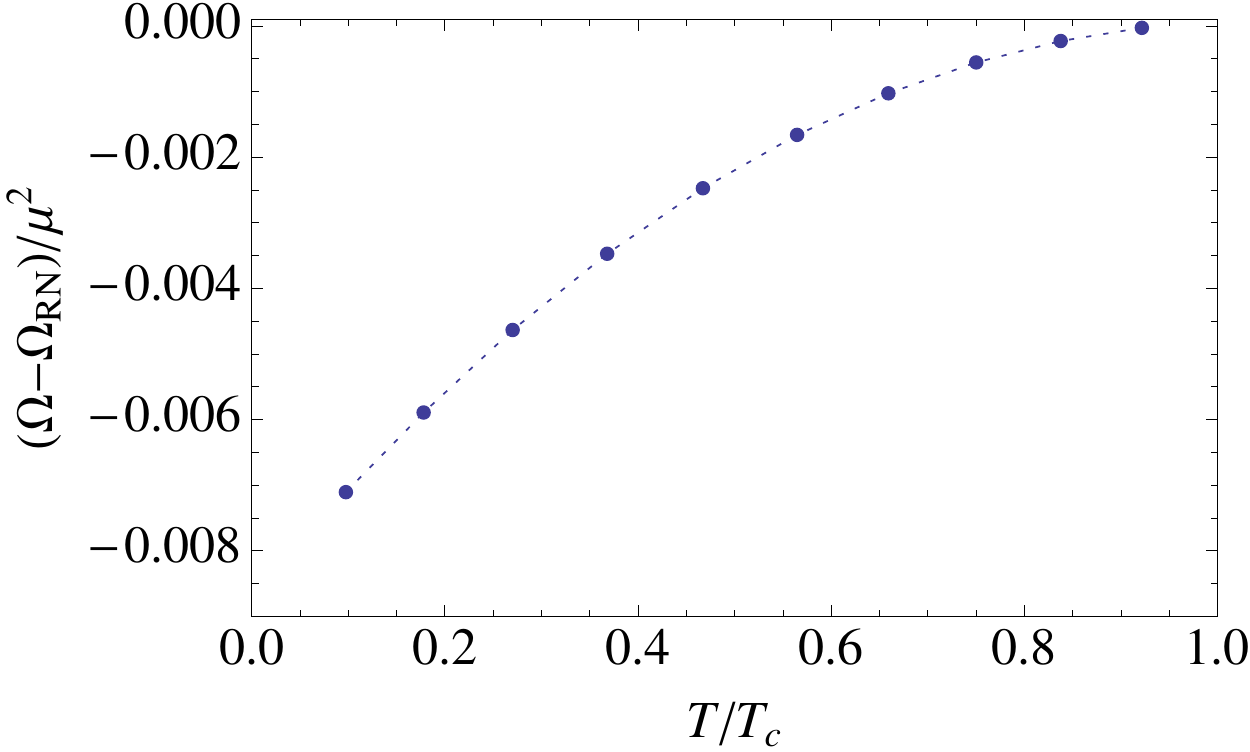}
    \caption[]{The grand free energy relative to the RN solution for $c_1=4.5$ and fixed $L\mu/4 = 2.08$. The grand free energies of the homogeneous and inhomogeneous phases are nearly degenerate, such that their maximal fractional difference is about 1\%.  }
\label{omega_c145}
\end{figure}

In this ensemble it is convenient to measure all quantities in units of the fixed chemical potential $\mu$. Then, after fixing $L$ from the critical mode appearing at the highest $T_c$ (see Fig.~\ref{linfig} and Table \ref{c1_Tc} in appendix \ref{lin_an}), we have that $\Omega/\mu^2$ is a function only of the dimensionless temperature $T/\mu$. In the fixed chemical potential ensemble for large enough axion coupling
we find a second order transition, where the inhomogeneous charge distribution starts dominating the thermodynamics immediately below the temperature at which the inhomogeneous instability develops. Near the critical temperature, the behavior of the grand free energy difference is consistent with $(\Omega-\Omega_{RN})/\mu^2\propto(1-T/T_c)^2$, while the entropy difference goes as $(S-S_{RN})/\mu\propto T/T_c-1$. This is as expected from a second order transition. As can be seen in Fig.~\ref{omegadiffs} and Fig.~\ref{omega_c145}, we find this second order transition for a range of lengths, $L$, and for a variety of values of the axion coupling $c_1$. 
With the current accuracy of our numerical procedure, we find it increasingly difficult to resolve the order of the phase transition for smaller values of $c_1$. In fact, for $c_1 = 4.5$ the grand free energies of the homogeneous and inhomogeneous phases are nearly degenerate but still allow us to determine the phase transition as second order.
 It would be interesting to see if the phase transition remains of second order or changes to the first order for smaller values of the axion coupling.


\begin{figure}
\center
    \includegraphics[width=15cm]{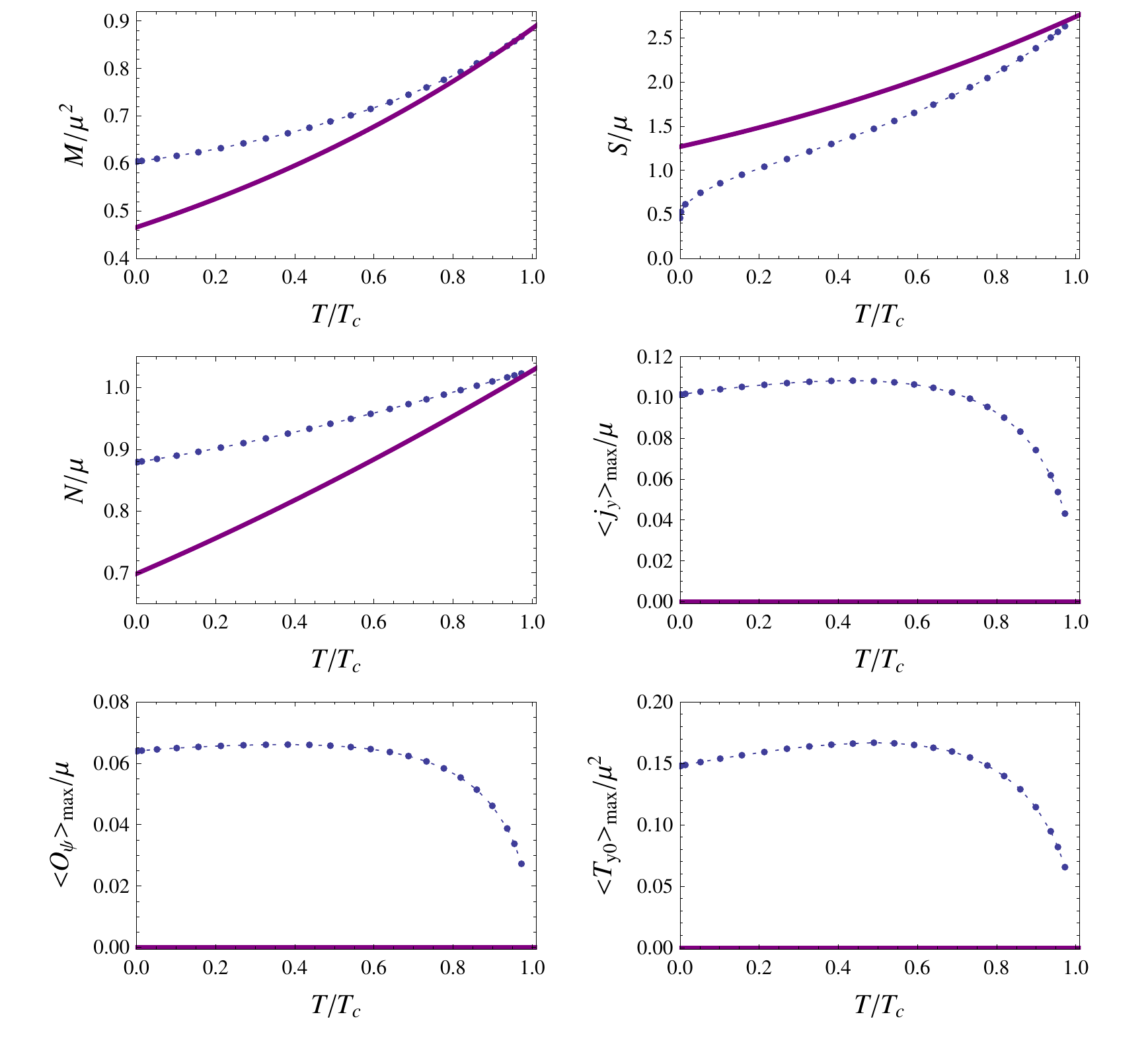}
    \caption[]{The observables in the grand canonical ensemble for $c_1=8$ and $L\mu/4=1.21$ (points with dotted line) plotted with the corresponding quantities for the RN black hole (solid line). Fitting the data near the critical point to the function $(1-T/T_c)^\alpha$, we find the approximate critical exponents $\alpha_{j_y}=0.40$, $\alpha_{T_y0}=0.41$ and $\alpha_{\mathcal{O}_\psi}=0.38$
with relative fitting error of about 10\%.}
\label{L121_obs}
\end{figure}

To examine the observables in the striped phase further, we focus on $c_1=8$ and the corresponding dominant critical mode, $L\mu/4\simeq1.21$, and consider solutions for the temperatures $0.00016 \lesssim T/T_c \lesssim 0.96$. Various quantities are plotted with the corresponding homogeneous results in Fig.~\ref{L121_obs}. Along this branch of solutions, the mass of the stripes is more than the RN solution and the entropy is always less. We plot the maximum of the boundary current density $\langle j_y\rangle$, momentum density $\langle T_{y0}\rangle$ and pseudoscalar operator vev $\langle \mathcal{O}_\psi\rangle$. Fitting the data near the critical point to the function $(1-T/T_c)^\alpha$, we find the approximate critical exponents $\alpha_{j_y}=0.40$, $\alpha_{T_y0}=0.41$ and $\alpha_{\mathcal{O}_\psi}=0.38$ with relative fitting error of about 10\%.  

We find evidence that the entropy of the striped black branes does not tend to zero in the small temperature limit, see Fig.~\ref{L121_obs}. This is further supported by the behavior of the transverse size of the horizon (\ref{ry}). Here the bulge seems to contract 
at a much slower rate than the neck, which evidently shrinks to zero size in the limit $T\rightarrow 0$. However, strictly speaking, this conclusion is based on extrapolation of the finite temperature data to $T=0$. Checking whether the entropy asymptotes to a finite value or goes to zero in this limit, as suggested in \cite{Withers:2013}, will require further investigation with a method of higher numerical accuracy. 

\subsubsection{The canonical ensemble}

To study the system in the canonical ensemble we fix the temperature, total charge and length of the system. This describes the physical situation in which the system is immersed in a heat bath consisting of uncharged particles. The relevant thermodynamic potential in this ensemble is the free energy  density 
\be F(T,N,L)= M-TS. \ee 
If we measure all quantities in units of the fixed charge $N$, then, again, the free energy $F/N^2$ is only a function of the dimensionless temperature $T/N$.

To solve our system with a fixed charge, we would need to fix the integral in $x$ of the coefficient of the $1/r$ term in the asymptotic expansion of the gauge field $A_t$. Numerically, it is much easier to fix the chemical potential, as this gives a Dirichlet condition on $A_t$ at the boundary. In the grand canonical ensemble, we solved for one-parameter families of solutions at fixed $L\mu$, labelled by the dimensionless temperature $T/\mu$. Equivalently, in the $(L\mu,T/\mu)$ plane, we solve along the line of fixed $L\mu$. Translated to the situation in which we measure quantities in terms of the charge density $N$, these solutions become one-parameter families of solutions with varying $LN$, or a curve in the $(LN,T/N)$ plane with $LN$ a function of $T/N$. By varying the length $L\mu$ (or solving with $\mu=1$ and varying $L$), we can find a collection of solutions that intersect the desired fixed $LN$ line. By interpolating these solutions and evaluating the interpolants at fixed $LN$, we can study the stripes in the canonical ensemble.

\begin{figure}
\center
    \includegraphics[width=10cm]{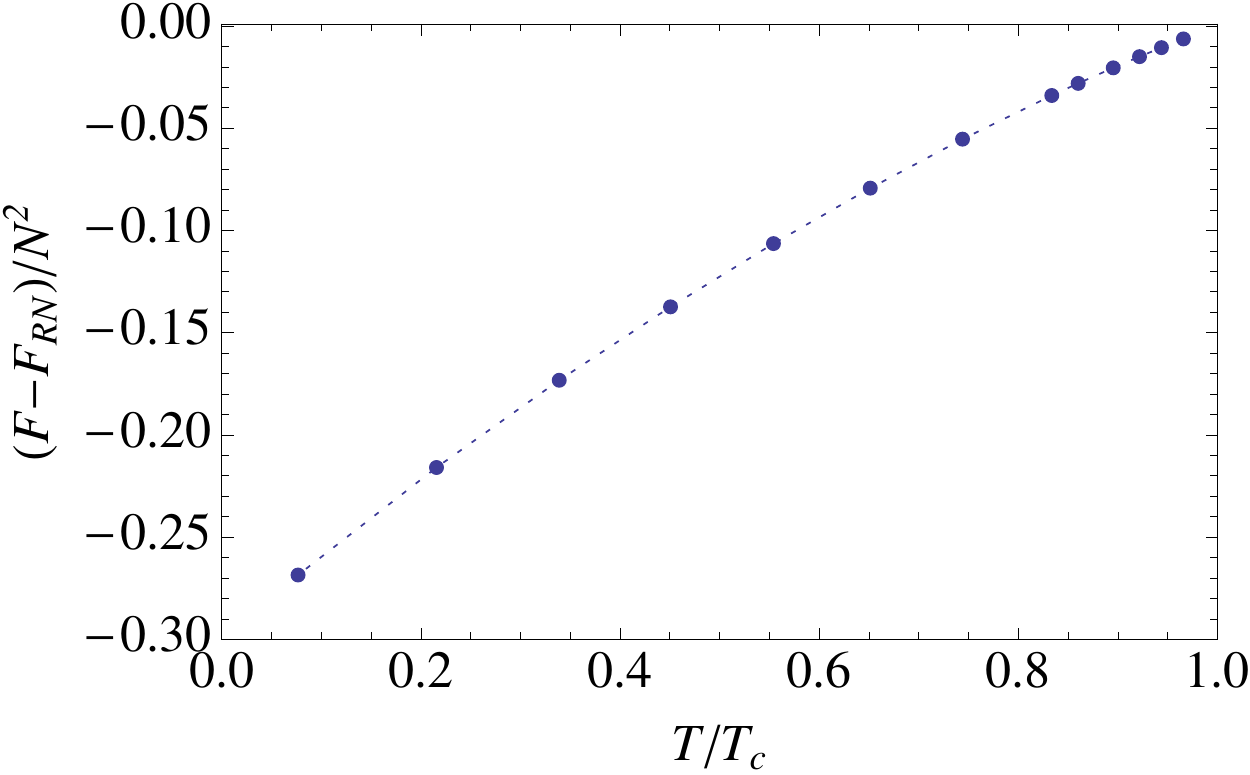}
    \caption[]{The difference in canonical free energy, at $c_1=8$ and fixed length $LN/4=1.25$, between the striped solution and the RN black hole. The striped solution dominates immediately below the critical temperature, signalling a second order phase transition.}
\label{Fdiff}
\end{figure}

In this ensemble we find a similar second order transition, in which the inhomogeneous solution dominates the thermodynamics below the critical temperature (Fig.~\ref{Fdiff}). The scaling of the relative free energy density slightly below the critical temperature appears to fit a linear scaling, however using more points at lower temperatures in the fit increases the critical exponent towards a quadratic scaling, as expected in a second order transition. 

\subsubsection{The microcanonical ensemble}

\begin{figure}
\center
    \includegraphics[width=10cm]{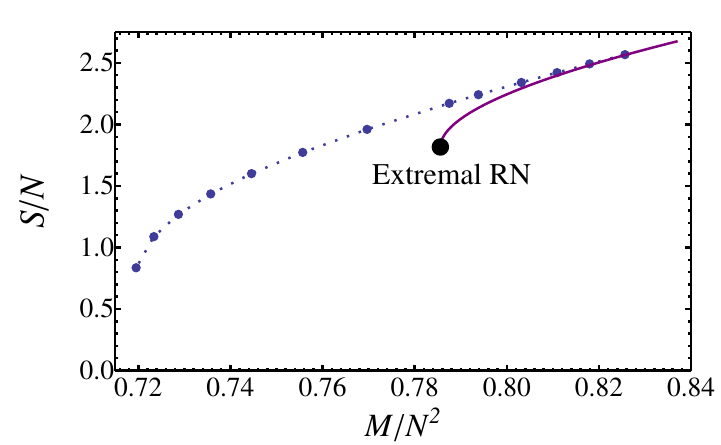}
    \caption[]{The entropy of the inhomogeneous solution for $c_1=8$ (points with dotted line) and of the RN solution (solid line). Below the critical temperature, the striped solution has higher entropy than the RN. The RN branch terminates at the extremal RN black hole, while the striped solution persists to smaller energies.}
\label{entropy_micro}
\end{figure}

The microcanonical ensemble describes an isolated system in which all conserved charges (in this case the mass and the charge) are fixed. This ensemble describes the physical situation relevant to the study of the real time dynamics of an isolated black brane at fixed length. In this case, the state that maximizes entropy is the dominant solution. As shown in Fig.~\ref{entropy_micro}, we find that the entropy of our inhomogeneous solutions is always greater than that of the RN black hole of the same mass. Furthermore, the mass of the inhomogeneous solutions is always smaller than that of the critical RN black hole. Therefore, at fixed $LN$, the unstable RN black holes below critical temperature are expected to decay smoothly to our inhomogeneous solution.

\subsubsection{Fixing the tension}
Alternatively, one could attempt to compare solutions with different values of $L$. The meaningful comparison is in an ensemble fixing the tension $\tau_x$. For example, one could compare the Legendre transformed grand free energy \be G(T,\mu,\tau_x)= M-TS-\mu N-\tau_x L \ee where the additional terms comes from boundary terms in the action rendering the new variational problem (fixing $\tau_x$) well-defined. The candidate saddle points are the solutions we find with various periodicities $L$, and their relative importance in the thermodynamic limit is determined by $G(T,\mu,\tau_x)$. In particular the solution which is thermodynamically dominant depends on the value of $\tau_x$ we hold fixed. In this study we concentrate on the thermodynamics in the fixed $L$ ensemble and we leave the study of the fixed $\tau_x$ ensemble to future work.

%===========================================================%
\section{Thermodynamics for the infinite system}
\label{sec_thermo_infinite}
In this section we lift the assumption of the finite extent of the system in the  $x$-direction and consider the thermodynamics of the
formation of the stripes below the critical temperature. For the infinite system we can define densities of thermodynamic quantities along $x$: \be
m=\frac{M}{L}, \;\;\; s=\frac{S}{L},\;\;\; n=\frac{N}{L}. \label{inf_dens} \ee In terms of these, the first law for the system becomes \be dm=Tds+\mu dn \label{inf_first}\ee and the conformal identity is \be 3m=2(Ts+\mu n). \label{inf_ward}\ee

\begin{figure}
\center
    \hspace*{2cm}
    \includegraphics[width=12cm]{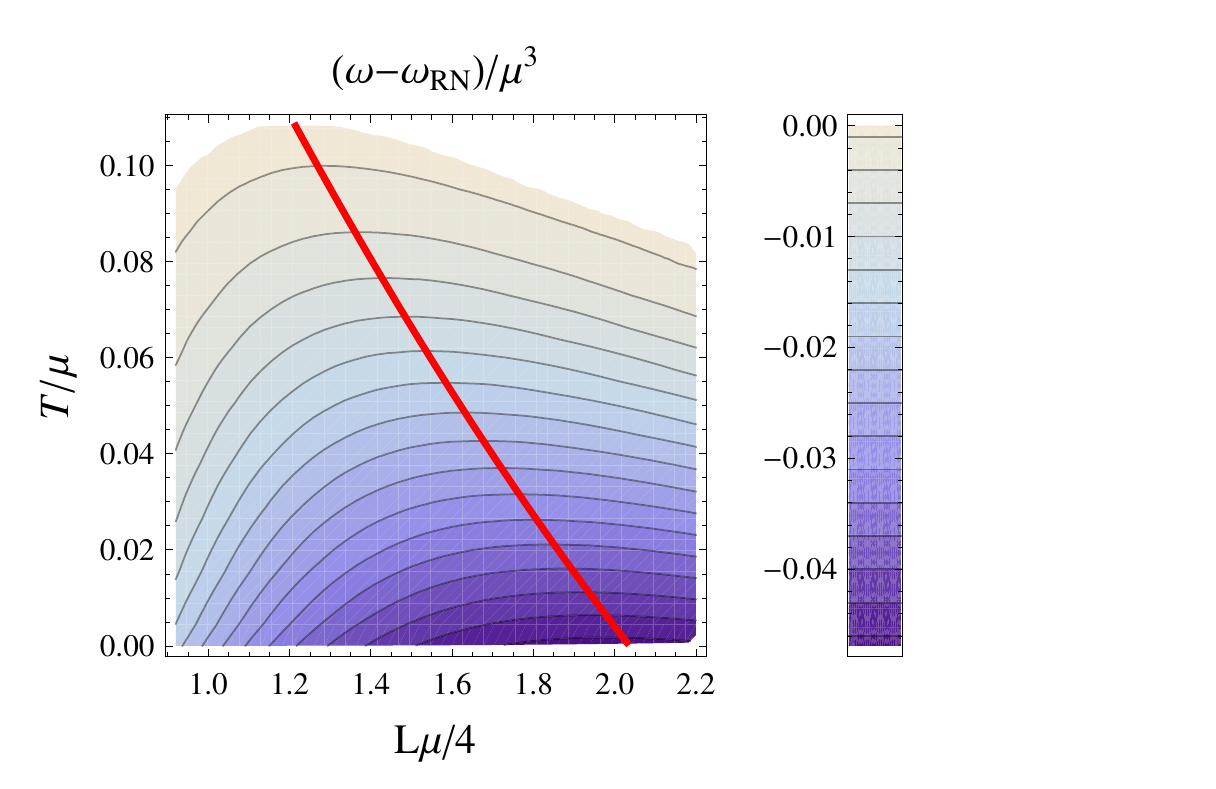}
    \caption[]{Action density for $c_1=8$ system relative to the RN solution. The red line denotes the approximate line of minimum free energy.}
\label{omega_density}
\end{figure}

In the infinite system, we compare stripes of different lengths, at fixed $T/\mu$, to each other and to the homogeneous solution. The solution that dominates the thermodynamics is the one with the smallest free energy density $\omega$, where \be \omega= m-Ts-\mu n. \ee This comparison is shown in Fig.~\ref{omega_density} for $c_1=8$, where we see that the free energy density of the stripes is negative relative to the RN black hole, indicating that the striped phase is preferred at every temperature below the critical temperature.\footnote{In appendix \ref{appx_inf}, we describe the generation of Fig.~\ref{omega_density}.} Very close to the critical temperature, the dominant stripe is that with the critical wavelength $k_c$. As we lower the temperature, the minimum of the free energy density traces out a curve in the $(L\mu,T)$ plane, and the dominant stripe width increases to $L\mu/4\approx2$.

One can also study the observables of the system along this line of minimum free energy density. The results are qualitatively similar to those for the fixed $L$ system (Fig.~\ref{L121_obs}). In particular, the free energy density scales as $(\omega-\omega_{RN})/\mu^3 \propto(1-T/T_c)^2$ near the critical point, indicating a second order transition in the infinite system as well.

%===========================================================%
\section*{Acknowledgements}
We are supported by a discovery grant from NSERC of Canada. ES was
partly supported by CITA National Fellowship.  We have
benefitted from conversations with Matt Choptuik, Aristomenis Donos,  Marcel Franz,  Jerome Gauntlett, Fernando Nogueira, Jorge Santos, Gordon Semenoff, Toby Wiseman and Mark van Raamsdonk.  The computer simulations were performed on WestGrid.

\appendix

\section{Asymptotic Charges}
\label{asympt_charges}

\subsection{Deriving the charges}

Since our ansatz is inhomogeneous and includes off-diagonal terms in the metric, and our action is not standard (in that it includes the axion coupling) we have re-derived the expressions for the charges and other observables in our geometry. In deriving the asymptotic charges of our spacetime, for the four dimensional Einstein-Maxwell-Higgs theory we discuss in the main text, we follow the covariant treatment of  \cite{Papadimitriou:2004ap, Papadimitriou:2005ii}. We refer the reader to those papers for details of the method used.

The bulk action has to be supplemented by boundary terms of two types. First, there are boundary terms needed to ensure that the variational problem is well-defined. Then there are counter-terms, terms depending only on the boundary values (leading non-normalizable modes) of fields on the cutoff surface, which are added to render the on-shell action and the conserved charges finite. Both kinds of  boundary terms are the standard ones for Einstein-Maxwell-Higgs theory; the additional axion coupling does not necessitate an additional boundary terms of either kind as long as the scalar mass satisfies $m^{2}<0$.

We find it convenient to study the first variation of the on-shell action, which always reduces to boundary terms. The expression for the regulated first variation of the on-shell action can be differentiated with respect to the boundary values of the bulk fields, to give finite expressions for the conserved charges. We write those expressions below in terms of the asymptotic expansion of the fields occurring in our ansatz, carefully taking into account the differences between our coordinate system and the standard Fefferman-Graham form of the asymptotic metric, which is used to derive the standard expressions in the literature.

Having explained our procedure, we now display the expressions for the observables used in the main text. We first assume the radial coordinate is in the standard Fefferman-Graham form, and then discuss additional terms arising from change of coordinate necessary to bring our asymptotic metric into the standard form.

For the scalar fields $\psi$, one can write asymptotically \be \psi(x,r)= \psi^{(0)}(x) r^{-\lambda_{-}} + \psi^{(1)}(x) r^{-\lambda_{+}} \ee with \be \lambda_{\pm}= \frac{3}{2} \pm \sqrt{\frac{9}{4}+m^{2}}.\ee We set $\psi^{(0)}(x) =0$ as part of our boundary conditions, then the coefficient $\psi^{(1)}(x)$ is the spatially modulated VEV of the scalar operator dual to $\psi$.

Similarly, the gauge field can be expanded near the boundary as \be A_{\mu}(x,r) = A_{\mu}^{(0)}(x) - \frac{A_{\mu}^{(1)}(x)}{r}. \ee The functions $A_{\mu}^{(1)}(x)$ correspond to the charge and current density of the boundary theory.

As for the boundary energy-momentum tensor, the expression is fairly simple in odd number of boundary dimensions, and we have checked that it is not modified by the matter action. With our normalization convention one can write
\be T_{ij} = 6g_{ij}^{{(3)}}, \ee
where the superscripts of the metric functions denote the order in the asymptotic expansion.

Since our metric ansatz is not of the Fefferman-Graham form, we need to perform a change of coordinate (in the $x,r$ plane, for  which we used the conformal ansatz) to put the metric is such a form. The details of the transformation are straightforward and the process results in the following shifts in the asymptotic metric quantities:
\be
\Delta g_{ij}^{(3)} = \frac{2}{3}  g(x),
\ee
for every $i,j$, where $g(x)$ is the leading asymptotic correction to the metric component $g_{rr}$. That is, at large $r$ that metric component becomes
\be
g_{rr}(r,x) \rightarrow  \frac {1}{2 r^2}+  \frac{g(x)}{r^5}.
\ee

Finally, since the metric becomes diagonal asymptotically, the
non-vanishing time components of the energy-momentum tensor $T_{tt}$
and $T_{yt}$ have a simple interpretation as energy and momentum
density, respectively. The conserved charges are given by integrating
those densities over a spatial slice.

\subsection{Explicit expressions for the charges}
\subsubsection{Homogeneous solution}
For reference, in this subsection we give the explicit expressions for the homogeneous RN solution in our conventions. The radius of the horizon is given in terms of the temperature by \be r_0=\frac{1}{6}\left(2\pi T+\sqrt{3\mu^2+4\pi^2T^2} \right).\ee The mass, entropy and charge of the RN solution of fixed length $L$ are \bea M_{RN}&=&\left(4r_0^3+\mu^2r_0 \right)L, \\ S_{RN}&=& 4\pi r_0^2 L, \\ N_{RN}&=&2r_0\mu L.\eea The corresponding densities in the infinite system are given by dividing through by $L$.

\subsubsection{Inhomogeneous solution}
\label{appx_inhomo_charges}
Here we list explicit expressions for the thermodynamic quantities in our system in terms of our solution ansatz. Conserved charges are given by integrating over the inhomogeneous direction. We define $f^{(3)}=-(4r_0^3+\mu^2r_0)/4$, the $1/r^3$ term from the function $f(r)$ (equation (\ref{frn})), and $X^{(3)}(x)$, for $X=\{R,S,T\}$, as the coefficient of the $1/r^3$ term of the corresponding metric function. The energy-momentum tensor yields the mass\footnote{See appendix \ref{appx_con} for details about the numerical process, including the definitions of the $\tilde{x}$ coordinate and $\xi(x)$. The functions $\{R,S,T\}$ are defined on the UV grid; they are analogous to $\{A,B,C\}$ in the original ansatz.} \be M = \int_0^L \langle T^{tt}(\tilde{x})\rangle d\tilde{x}  =4\int_0^{L} \xi(x)^2 (-f^{(3)} +5S^{(3)}(x)+3T^{(3)}(x))dx,\label{mass}\ee the tension in the $x$ direction \be \tau_x = - \int_0^L \langle T^{xx}(\tilde{x})\rangle d\tilde{x}  = 2 \int_0^{L}  \xi(x)^2 (f^{(3)} +6R^{(3)}(x)+ 4S^{(3)}(x)+6T^{(3)}(x))dx,\label{tension} \ee and the pressure in the $y$ direction \be P_y = \int_0^L \langle T^{yy}(\tilde{x})\rangle d\tilde{x}  = -2 \int_0^{L}\xi(x)^2 (f^{(3)} + 6R^{(3)}(x) + 10S^{(3)}(x))dx.\ee Now, expanding the equations of motion at the asymptotic boundary, we get the relation $R^{(3)}(x)+2S^{(3)}(x)+T^{(3)}(x)=0$. Using this, we see that $\langle T^{\mu\nu}(z) \rangle$ is traceless, as necessary. Conservation of the energy momentum tensor requires $\partial_x\tau_x=0$. This is related to the constraint equation (\ref{asymp_con}) and we explain our strategy to ensure it is satisfied in appendix \ref{appx_con}.
 
 The coefficient of the $1/r$ falloff of the gauge field gives the charge \be N = -2\int_0^{L} A_t^{(1)}(x).\ee At the horizon, we read the (constant) temperature as \be T=\frac{1}{8\pi r_0}(12r_0^2 - \mu^2)e^{-(B-A)|_{r=r_0}}\ee and the entropy is proportional to the area of the event horizon, given by \be S = 4\pi r_0^2 \int_0^{L/4} e^{(B(r_0,x)+C(r_0,x))}dx. \ee

\subsection{Consistency of the first laws}
\label{first_appx}
Here, we discuss the first laws for both the finite length stripe and the stripe on the infinite domain.
\subsubsection{Finite system}
In our system, as described above, we have unequal bulk stresses $\tau_x$\footnote{We define $\tau_x=-P_x$, where $P_x$ is the pressure in the $x$ direction. For our solutions, $\tau_x>0$.} and $P_y$. Then, if we have a rectangle of side lengths $(L,L_y)$, the work done by the expansion or compression of this region will differ depending on which direction the stress is in. The usual $-PdV$ term in the first law is replaced and we have \be d\hat{M}=Td\hat{S}+\mu d\hat{N} + \tau_x L_ydL - P_y L dL_y, \ee where the hatted variables represent thermodynamic quantities integrated over the entire system. Defining densities (in the trivial $y$-direction) by \be M=\frac{\hat{M}}{L_y},\;\;\; S=\frac{\hat{S}}{L_y},\;\;\; N=\frac{\hat{N}}{L_y},\ee we can write the first law as \be dM=TdS+\mu dN + \tau_x dL +\frac{dL_y}{L_y}(-M+TS+\mu N - P_yL). \ee Tracelessness of the energy-momentum tensor implies $M=L(P_x+P_y)$, so that the term proportional to $dL_y$ can be rewritten as the conformal identity (\ref{ward}), which disappears for a conformal system described by the first law (\ref{first}). Therefore, the first law (\ref{first}) and the conformal identity (\ref{ward}) are consistent.

\subsubsection{Infinite system}
For the infinite system, we define densities in both the $x$ and $y$ directions as equation (\ref{inf_dens}). Under the scaling symmetry (\ref{scaling_symmetry}), these scale as \be m\rightarrow\lambda^3 m, \;\;\; s\rightarrow\lambda^2 s, \;\;\; n\rightarrow\lambda^2 n. \ee Using the first law (\ref{inf_first}), we derive the conformal identity (\ref{inf_ward}). Again, we can see this from the first law for the system with integrated charges. Plugging the densities $m,s,n$ into the first law of the finite length system (\ref{first}), we arrive at \be dm=Tds+\mu dn +\frac{dL}{L}(-m+Ts+\mu s +\tau_x). \ee Using the conformal identity of the finite length system (\ref{ward}), we see that the term proportional to $dL$ is just the conformal identity for the infinite system, which is satisfied for a system described by (\ref{inf_first}).

%========================================================
\section{Further details about the numerics}
\label{appendixB}

\subsection{The linearized analysis}
\label{lin_an}
Following \cite{Donos:2011bh}, we look for static normalizable modes around the Reissner-Nordstrom background. We consider the fluctuation\footnote{Regularity at the black hole horizon enforces that $\delta g_{ty}(r_0)=0$.} \bea \delta g_{ty} &=& \lambda\left( \frac{(r-r_0)}{r}w(r) \sin (kx)\right), \nn\\ \delta A_y&=&\lambda(a(r) \sin(kx)),\nn \\ \delta\psi&=&\lambda (\phi(r) \cos(kx)),\eea where $\lambda$ is a small parameter in which we can expand the equations. Putting this ansatz into (\ref{eqA}) - (\ref{eqAy}) and expanding to linear order in $\lambda$, we arrive at the linearized system \be w''(r)-\frac{{r_0} a'(r)}{r^3(r-{r_0})}+\frac{(4 r-2 {r_0}) w'(r)}{r(r- {r_0})}+\frac{w(r) \left(2 {r_0}\left(4 r^3+4 r^2 {r_0}+4 r{r_0}^2-{r_0}\right)-k^2 r^2\right)}{r^2\left(4 r^4-r \left(4{r_0}^3+{r_0}\right)+{r_0}^2\right)}=0,\nn\ee 
\bea &&a''(r)+\frac{\left(8 r^4+r \left(4{r_0}^3+{r_0}\right)-2 {r_0}^2\right)a'(r)}{r \left(4 r^4-r \left(4{r_0}^3+{r_0}\right)+{r_0}^2\right)}-\nn\\ && \frac{k^2 a(r)}{4 r^4-r \left(4{r_0}^3+{r_0}\right)+{r_0}^2} +\frac{{c_1} k {r_0} \phi(r)}{\sqrt{3} \left(4 r^4-r\left(4{r_0}^3+{r_0}\right)+{r_0}^2\right)}-\nn\\ &&\frac{4 r {r_0} w'(r)}{4 r^3+4 r^2 {r_0}+4 r{r_0}^2-{r_0}}-\frac{4 {r_0}^2 w(r)}{4r^4-r \left(4{r_0}^3+{r_0}\right)+{r_0}^2}=0, \eea 
\bea && \phi''(r)+\frac{{c_1} k {r_0} a(r)}{2 \sqrt{3} r^2\left(4 r^4-r \left(4{r_0}^3+{r_0}\right)+{r_0}^2\right)}- \nn\\ &&\frac{\phi(r) \left(k^2+2 m^2 r^2\right)}{4 r^4-r \left(4{r_0}^3+{r_0}\right)+{r_0}^2} -\frac{\left(-16 r^3+4 {r_0}^3+{r_0}\right) \phi'(r)}{4r^4-r \left(4{r_0}^3+{r_0}\right)+{r_0}^2}=0.\nn\eea
Fixing the scalar field mass as $m^2=-4$,
 there are three parameters in these equations: the temperature of the black brane $T_0$ (equivalently the location of the horizon $r_0$), the wavenumber $k$, and the strength of the axion coupling $c_1$. In this analysis, we will choose $c_1$ and $k$ and then use a shooting method to find the $T_0$ at which normalizable modes appear.
 
Due to the linearity of the equations, the scale of our solutions is arbitrary. We use this to fix a Dirichlet condition on $w$ at the horizon. Changing coordinates to $\rho=\sqrt{r^2-r_0^2}$,  and expanding the equations near $\rho=0$ gives regularity conditions on the fluctuations at the horizon in terms of Neumann boundary conditions. Our horizon boundary conditions are then
 \be w(\rho)|_{\rho=0}=1,\;\;\;\;\;\; w'(\rho)|_{\rho=0}=a'(\rho)|_{\rho=0}=\phi'(\rho)|_{\rho=0}=0,\ee 
Namely, that the fields are quadratic in $\rho$ near the horizon. 
In order to search for normalizable modes, we set the sources in the field theory to zero by imposing leading order fall-off conditions near the AdS boundary: 
\be w(\rho)=\frac{w_3}{\rho^3}+\dots, \;\;\; a(\rho)=\frac{a_1}{\rho}+\dots,\;\;\; \phi(\rho)=\frac{\phi_2}{\rho^2}+\dots.\ee 
In practice, after fixing $c_1$ and $k$, we use $T_0$ as a shooting parameter to find the solution with the correct $w$ fall-off and the corresponding critical temperature $T_c$.

\begin{figure}
\center
    \includegraphics[width=10cm]{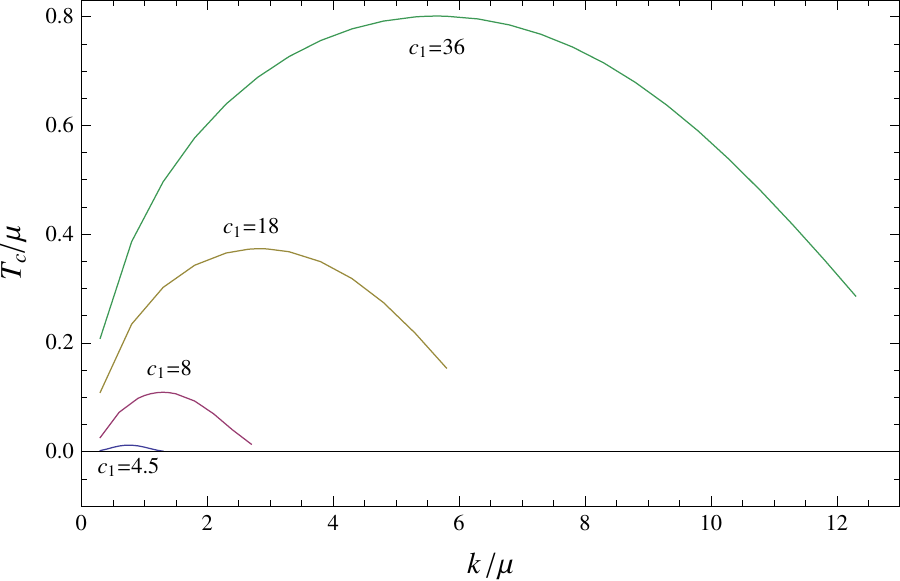}
    \caption[]{The critical temperatures at which the Reissner Nordstrom black brane becomes unstable, for varying axion coupling $c_1$. As the strength of the axion coupling increases, the size of the unstable region (the area under the critical temperature curve) also increases.}
\label{linfig}
\end{figure}

For each $c_1$, we find a range of unstable momenta. By adjusting the strength of the axion coupling, one can find a large variation in the size of this unstable region in the $(k/\mu,T_0/\mu)$ plane (see Fig.~\ref{linfig}). The relationship between $c_1$ and the maximum critical temperature is well fit by $T_{c}^{max}(c_1)/\mu=0.025c_1-0.091$. The wavenumbers for the dominant critical modes, corresponding to $T_c^{max}(c_1)$, for select $c_1$ are found in Table \ref{c1_Tc}.

\begin{table}[h!]
\centering
\begin{tabular}{| c | c | c | c |}
\hline
$c_1$ & $T_{c}^{max}/\mu$ & $k_c/\mu$ & $L\mu/4=\pi/2k_c$\\
\hline
4.5 & 0.012 & 0.75 & 2.08 \\
5.5 & 0.037 & 0.92 & 1.71 \\
8 & 0.11 & 1.3 & 1.21 \\
18 & 0.37 & 2.85 & 0.55 \\
36 & 0.80 & 5.65 & 0.28 \\
\hline
\end{tabular}
\caption{The maximum critical temperatures and corresponding critical wavenumbers for varying $c_1$.}
\label{c1_Tc}
\end{table}

%
 
%========================================================
\subsection{The equations of motion}
\label{eom}
For completeness, here we present the equations of motion derived from the Lagrangian (\ref{Lgrn}). The Einstein equations in our case are four second order elliptic equations for the metric components and two constraint equations. For the compactness of the expressions, we define
\be
\ohat U \cdot \ohat V = \dr U\dr V + \frac{1}{4r^4f}\dx U\dx V, \;\;\;\; \ohat^2U=\dr^2U+\frac{1}{4r^4f}\dx^2U.
\ee
The four elliptic equations, formed from combinations of $G^t_t-T^t_t=0$, $G^t_y-T^t_y=0$, $G^y_y-T^y_y=0$, and $G^r_r+G^x_x-(T^r_r+T^x_x)=0$, then take the form

\bea
&&\ohat^2A + (\ohat A)^2  +\ohat A\cdot\ohat C -\frac{e^{-2A+2C}}{2f}(\ohat W)^2 -\frac{e^{-2A}}{4r^2f}(\ohat A_t)^2 \nn
\\ &&-\frac{1}{4r^2}\left( \frac{e^{-2A}W^2}{f}+ e^{-2C}  \right)(\ohat A_y)^2 -\frac{e^{-2A}W}{2r^2f}\ohat A_t\cdot\ohat A_y   +\left(\frac{5 }{r}+\frac{3 f'}{2 f}\right) \dr A \nn
\\ &&+\left(\frac{1}{r}+\frac{f'}{2 f}\right)\dr C  +\frac{3}{r^2}-\frac{3 e^{2B}}{r^2f}+\frac{e^{2B} m^2 \psi ^2}{4  r^2 f}+\frac{3 f'}{r f}+\frac{f''}{2 f}
= 0,\label{eqA}
\eea
\bea
&&\ohat^2 B +\half (\ohat\psi)^2 -\frac{e^{-2A+2C}}{4f}(\ohat W)^2 - \ohat A\cdot\ohat C-\frac{1}{r}\dr A \nn
\\ &&+\left( \frac{2}{r}+\frac{f'}{2 f}\right)\dr B-\left(\frac{1}{r}+\frac{f'}{2 f}\right)\dr C= 0, \label{eqB}
\eea
\bea
&&\ohat^2 C + (\ohat C)^2 + \ohat A\cdot\ohat C + \frac{e^{-2A+2C}}{2f}(\ohat W)^2 + \frac{e^{-2A}}{4r^2f} (\ohat A_t)^2 \nn
\\ &&+\frac{1}{4r^2}\left( \frac{e^{-2A}W^2}{f}+ e^{-2C} \right) (\ohat A_y)^2 + \frac{e^{-2A}W}{2r^2f} \ohat A_t \cdot\ohat A_y \nn
\\ &&+\frac{1}{r} \dr A+\left( \frac{5}{r}+\frac{f' }{f} \right) \dr C+\frac{3}{r^2}-\frac{3 e^{2B}}{r^2 f}+\frac{e^{2B} m^2 \psi ^2}{4 r^2 f}+\frac{f'}{ r f}= 0, \label{eqC}
\eea
and
\bea
&&\ohat^2W- \ohat A\cdot\ohat W + 3\ohat C\cdot\ohat W - \frac{e^{-2C}W}{r^2}(\ohat A_y)^2 \nn
\\ &&- \frac{e^{-2C}}{r^2} \ohat A_t \cdot\ohat A_y +\frac{4}{r} \dr W= 0. \label{eqW}
\eea

The matter field equations are
\bea
&&\ohat^2 \psi + \ohat A\cdot\ohat\psi + \ohat C\cdot\ohat\psi + \frac{c_1 e^{-A-C}}{8\sqrt3 r^4f}\left( \dr A_t \dx A_y - \dx A_t \dr A_y\right) \nn
\\  && +\left(\frac{4}{r}+\frac{f'}{f} \right)\dr\psi-\frac{e^{2 B} m^2 \psi }{2 r^2 f}=0, \label{eqpsi}
\eea
\bea
&&\ohat^2 A_t -\ohat A\cdot\ohat A_t + \ohat C\cdot\ohat A_t + \frac{e^{-2A+2C}W}{f} \ohat W\cdot\ohat A_t + \ohat W\cdot\ohat A_y  \nn
\\ &&+ 2W \ohat C\cdot\ohat A_y - 2W \ohat A\cdot\ohat A_y +\frac{e^{-2A+2C}W^2}{f} \ohat W\cdot\ohat A_y  \nn
\\ &&+\frac{c_1}{4\sqrt3 r^2} \left( e^{A-C} - \frac{ e^{-A+C}W^2}{f}  \right) (\dr \psi \dx A_y - \dx\psi \dr A_y) \nn
\\ &&- \frac{c_1 e^{-A+C}W}{4\sqrt3 r^2 f} ( \dr\psi \dx A_t - \dx\psi \dr A_t) +\frac{2}{r}\dr A_t -\frac{W f'}{f}\dr A_y=0, \label{eqAt}
\eea
and
\bea
&&\ohat^2 A_y + \ohat A\cdot\ohat A_y - \ohat C\cdot\ohat A_y - \frac{e^{-2A+2C}W}{f}\ohat W\cdot\ohat A_y -\frac{e^{-2A+2C}}{f}\ohat W\cdot\ohat A_t \nn
\\ &&+ \frac{c_1e^{-A+C}}{4\sqrt3 r^2 f} (\dr\psi \dx A_t - \dx\psi\dr A_t) + \frac{c_1 e^{-A+C}W}{4\sqrt3 r^2 f}(\dr\psi\dx A_y - \dx\psi\dr A_y) \nn
\\ &&+\left( \frac{2}{r}+\frac{f'}{f}\right) \dr A_y= 0. \label{eqAy}
\eea 

Finally, the constraint equations are
\bea
&&   -2 e^{-2 B} f r^2 (\dx \dr A+ \dx\dr C)+2e^{-2 B} f r^2 \dr A\left(\dx B- \dx A\right)\nn\\
  && +2 e^{-2 B} f r^2\left(\dx A+\dx C\right) \dr B+2 e^{-2 B} f r^2\left(\dx B-\dx C\right)\dr C -e^{-2 B} r^2f' \dx A\nn\\
  &&    +e^{-2 B} \left(r^2 f'+4 fr\right)\dx B+e^{-2 (A+B)}\left(\dx A_t +W\dx A_y\right)\left(\dr A_t+W\dr A_y\right)\nn\\
  &&+r^2e^{-2 (A+B-C)}\dx W\dr W
  -f e^{-2 (B+C)}\dx A_y \dr A_y 
   -2e^{-2 B} f r^2 \dx\psi\dr\psi=0\nn\\
\eea
and
\bea
&&\dr^2A+\dr^2C-\frac{1}{4 fr^4}(\dx^2A+\dx^2C)+ \left(1-\frac{1}{4 fr^4}\right) \left(\dr A\right)^2+\left(1-\frac{1}{4 fr^4}\right) \left(\dr C\right)^2\nn\\
&&+\frac{1}{2 fr^4}\left(\dx A+\dx C\right)\dx B-2 \left(\dr A+\dr C\right)\dr B+ \left(\frac{3 f'}{2f}+\frac{2}{r}\right)\dr A-\left(\frac{f'}{f}+\frac{4}{r}\right)\dr B\nn\\
  && + \left(\frac{f'}{2f}+\frac{2}{r}\right)\dr C+\frac{e^{-2 A}}{8 f^2r^6}\left(\dx A_t+W\dx A_y\right)^2-\frac{e^{-2 A}}{2 fr^2}\left(\dr A_t+W\dr A_y\right)^2\nn\\
  && -\frac{e^{-2 (A-C)}}{2 f}\left(\left(\dr W\right)^2-\frac{1}{4 fr^4}\left(\dx W\right)^2\right)+\frac{e^{-2 C}}{2r^2}\left(\left(\dr A_y\right)^2-\frac{1}{4 f r^4}\left(\dx A_y\right)^2\right)\nn\\
 && +\left(\dr\psi\right)^2-\frac{1}{4 fr^4}\left(\dx\psi\right)^2+\frac{f''}{2f}+\frac{2 f'}{f r}=0.
\eea

\subsection{Constraints}
\label{appx_con}

The constraint equations, $G^r_x-T^r_x=0$ and $G^r_r-G^x_x-(T^r_r-T^x_x)=0$, are the non-trivial Einstein equations that are not part of the system of second-order elliptic equations that we numerically solve. As discussed in \S\ref{equations}, the weighted constraints can be shown to solve Laplace equations on the domain. If we satisfy one of the constraints on all boundaries and the other at one point, they will be satisfied everywhere. At the black hole horizon, we choose to impose $r^2 \sqrt{f} \rootg (G^r_r-G^x_x-(T^r_r-T^x_x))=0$ at the point $(\rho,x)=(0,0)$ and $\rootg (G^r_x-T^r_x)=0$ across the horizon. Since we use periodic boundary conditions in the inhomogeneous direction, the boundaries at $x=0$ and $x=x_{max}$ are trivial if $\rootg (G^r_x-T^r_x)=0$ at the horizon and the conformal boundary. Then, we are left with the task of satisfying $\rootg (G^r_x-T^r_x)=0$ at the boundary.

In \S\ref{equations}, we found the asymptotic expansion of this constraint as \be G^r_x-T^r_x\propto\frac{3\dx A^{(3)}(x)+2\dx B^{(3)}(x)+3\dx C^{(3)}(x)}{r^2}+O(r^{-3}),\ee where $A^{(3)}(x),B^{(3)}(x)$ and $C^{(3)}(x)$ come from solving the elliptic equations. It appears that, within our problem, we do not have the ability to make the weighted constraint disappear. The key lies in an unfixed gauge symmetry in our original metric that is related to conformal transformations of the $(r,x)$ plane.\footnote{See \cite{Aharony:2005bm} for a discussion of the same issue in a different context.}
 Essentially, within our metric ansatz, we have the freedom to transform to any plane $(r',x')$ that is conformally related to $(r,x)$. Demanding that the weighted constraint $\rootg (G^r_x-T^r_x)$ vanishes at the conformal boundary uniquely identifies the correct coordinates $(\tilr,\tilx)$.

Our procedure is to split the domain at some intermediate radial value $\rho_{int}$. On the IR portion of the grid, $0<\rho<\rho_{int}$, the equations are as above. On the UV portion of the grid, $\rho_{int}<\rho<\rho_{cut}$, we use the coordinate freedom to select the correct asymptotic radial coordinate. We can write the metric in the UV as \bea ds^2=-2\tilr^2\tilde{f}(\tilr,\tilx)e^{2R}dt^2+e^{2S}\left(\frac{d\tilr^2}{2\tilr^2\tilde{f}(\tilr,\tilx)}+2\tilr^2d\tilx^2\right)+2\tilr^2e^{2T}(dy-Udt)^2,\eea where $\tilde{f}(\tilr,\tilx)\equiv f(r(\tilr,\tilx))$. Under a transformation in the $(\tilr,\tilx)$ plane such that $\tilr$ and $\tilx$ satisfy Cauchy-Riemann-like relations \be \frac{\partial\tilr(r,x)}{\d r} = \frac{\tilr(r,x)^2}{r^2}\frac{\d\tilx(r,x)}{\d x}, \;\;\;\;\;\; \frac{\d\tilx(r,x)}{\d r}=-\frac{1}{4r^2\tilr(r,x)^2f(r)}\frac{\d\tilr(r,x)}{\d x}, \ee the metric becomes \bea ds^2=-2\tilr(r,x)^2f(r)e^{2R}dt^2+e^{2S}|\nabla\tilr(r,x) |^2\left(\frac{dr^2}{2r^2f(r)}+2r^2dx^2\right)+2\tilr(r,x)^2e^{2T}(dy-Udt)^2 \label{UV_metric}\nn\\\eea with \be |\nabla\tilr(r,x) |^2 = \frac{r^2}{\tilr(r,x)^2}\left( \frac{\partial\tilr(r,x)}{\partial r} \right)^2+\frac{1}{4r^2\tilr(r,x)^2f(r)} \left( \frac{\partial\tilr(r,x)}{\partial x} \right)^2.\ee We now have an extra function $\tilr(r,x)$ in our system which we may use to satisfy the constraint and fix the residual gauge freedom, as we will now see. The Cauchy-Riemann-like conditions give the Laplace-like equation \be \frac{\d}{\d r}\left(\frac{r^2}{\tilr(r,x)^2}\frac{\d\tilr(r,x)}{\d r} \right)+\frac{\d}{\d x}\left( \frac{1}{4r^2\tilr(r,x)^2f(r)} \frac{\d\tilr(r,x)}{\d x}\right)=0. \label{tilr_eq}\ee We can solve this asymptotically, finding \be\tilr(r,x) =\xi(x)r+\frac{2\xi'(x)^2-\xi(x)\xi''(x)}{24\xi(x) r}+\dots, \ee where $\xi(x)$ is an arbitrary function that encodes the coordinate freedom we have.

Expanding the constraint asymptotically, we have \bea G^r_x-T^r_x&\propto&\frac{1}{r^2}\Big( 2(3\d_xR^{(3)}(x)+2\d_xS^{(3)}(x)+3\d_xT^{(3)}(x))\xi(x)\nn\\ &&+\;3(f^{(3)}+2R^{(3)}(x)-4S^{(3)}(x)+2T^{(3)}(x))\xi'(x)\Big)+O(r^{-3}),\nn\\ \label{asympt_con2} \eea where $X=X^{(3)}(x)/r^3+\dots$ asymptotically, for $X=\{R,S,T\}$. 
Demanding that the constraint (\ref{asympt_con2}) vanishes at the leading order yields a differential equation we can solve for $\xi(x)$, giving us a boundary condition for the function $\tilr(r,x)$, such that the weighted constraint will disappear at the conformal boundary. However, we have found that the code is unstable if we directly use this solution for $\xi(x)$. Instead of directly integrating the constraint, we use the freedom in $\xi(x)$ to fix the tension $\tau_x$ to be constant. This enforces the same effect on the tension as if we had used the explicit solution for $\xi(x)$ but is much more stable numerically. Below, we check that the constraints are suitably satisfied even though our boundary conditions do not exactly fix them. To this end, we set \be\xi(x)= \frac{K}{(f^{(3)} +6R^{(3)}(x)+ 4S^{(3)}(x)+6T^{(3)}(x))^{1/3}}.\label{xi}\ee Expanding the equations asymptotically gives the expression $R^{(3)}(x)+2S^{(3)}(x)+T^{(3)}(x)=0$; if this is satisfied on our solutions our definition of $\xi(x)$ coincides with that found by integrating the constraint (\ref{asympt_con2}).

The constant $K$ appearing in $\xi(x)$ sets the scale of the boundary theory. We use it to fix the length of the inhomogeneous direction in the field theory to be $L\mu/4$. The correct coordinate in the inhomogeneous direction of the field theory is $\tilx$. From the Cauchy-Riemann conditions, we can find the large $r$ expansion of $\tilx(r,x)$ as \be\tilx(r,x)=\int_0^x \frac{dx'}{\xi(x')}+\frac{\xi'(x)}{8\xi(x)^2r^2}+\dots. \ee Integrating to find the proper length of one cycle in the boundary, we solve for $K$ at leading order in $r$ to find \be K=\frac{4}{L}\int_0^{L/4}(f^{(3)} +6R^{(3)}(x)+ 4S^{(3)}(x)+6T^{(3)}(x))^{1/3}dx'.\ee When integrating the charges over the inhomogeneous direction in the field theory, one must remember to integrate over the correct coordinate, $d\tilx=dx/\xi(x)$.

Our corrected numerical procedure is as follows. On the IR grid, we solve the elliptic equations (\ref{eqA}) - (\ref{eqAy}) for the metric functions $A,B,C$ and $W$. On the UV grid, we solve the equivalent elliptic equations from the metric (\ref{UV_metric}) in the variables $R,S,T$ and $U$ plus equation (\ref{tilr_eq}) for the new field $\tilr(r,x)$. At the horizon, we enforce the boundary conditions discussed in \S \ref{equations}. At the interface $\rho=\rhoi$, we impose matching conditions on the four metric functions and that $\tilr(\rhoi,x)=r(\rhoi)$. Asymptotically, $R,S,T$ and $U$ all fall off as $1/\tilr^3$. To set boundary conditions on $\tilr$, we notice that \be \d_r\tilr(r,x)+\frac{\tilr(r,x)}{r}=2\xi(x)+O\left(\frac{1}{r^3}\right).\ee We truncate this expression at $O(r^{-2})$ and finite difference to find an update procedure for $\tilr(\rho_{cut},x)$. This boundary condition is updated iteratively as the functions $R,S,T$ are updated in our solving procedure such that once we find a solution with small residuals we can be sure that the tension is constant and the constraint is satisfied.

\subsection{Generating the action density plot}
\label{appx_inf}

\begin{figure}[t]
\center
    \includegraphics[width=10cm]{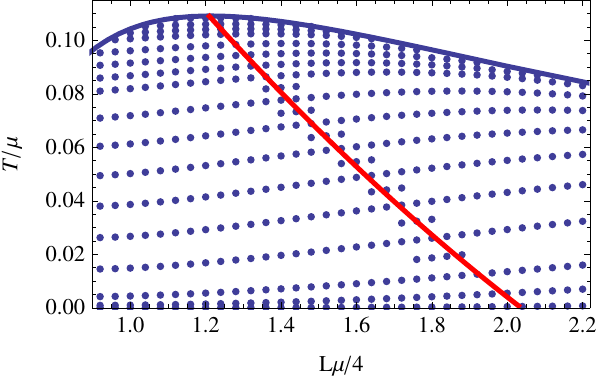}
    \caption[]{The data underlying Fig.~\ref{omega_density}. The points represent solutions we computed. These were interpolated to find the free energy density over the domain. The solid blue line is the edge of the unstable region and the thick red line is the approximate line of minimum free energy density.}
\label{dens_grid}
\end{figure}

To generate the relative action density plot, Fig.~\ref{omega_density}, we find the solutions on a grid of lengths $L$ and temperatures $T_0$, as shown in Fig.~\ref{dens_grid}. By interpolating these solutions on the domain, we can map the thermodynamic quantities across the unstable region and determine the approximate line of minimum free energy, or the dominant solution in the infinite size system.

\subsection{Convergence and independence of numerical parameters}

\subsubsection{Performance of the method and convergence of physical data}
\label{appx_conv}
As discussed above, to solve the equations numerically, we use a second order finite differencing approximation (FDA) before using a point-wise Gauss-Seidel relaxation method on the resulting algebraic equations. The method, including the UV procedure described above, performs well for this system.

\begin{figure}[t]
\center
    \includegraphics[width=10cm]{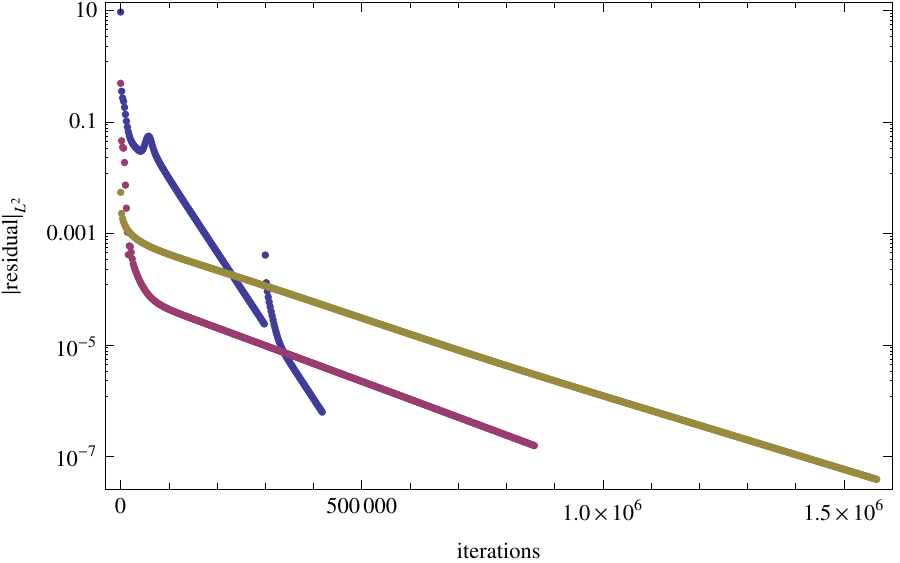}
    \caption[]{The behavior of the $L^2$ norm of the residual during the relaxation iterations for $c_1=8$, $T_0=0.04$ and $L\mu/4=0.75$. From top to bottom (at the left of the plot) the grid spacing is $d\rho, dx={0.04,0.02,0.01}$. The UV procedure is unstable unless the solution is close enough to correct solution. For grid spacing $d\rho, dx={0.04}$, the UV procedure was activated after $3\times 10^5$ iterations while for the others, the initial guess was taken to be a solution with slightly different parameters such that the UV procedure could be used immediately.}
\label{resids}
\end{figure}

\begin{figure}
\center
    \includegraphics[width=10cm]{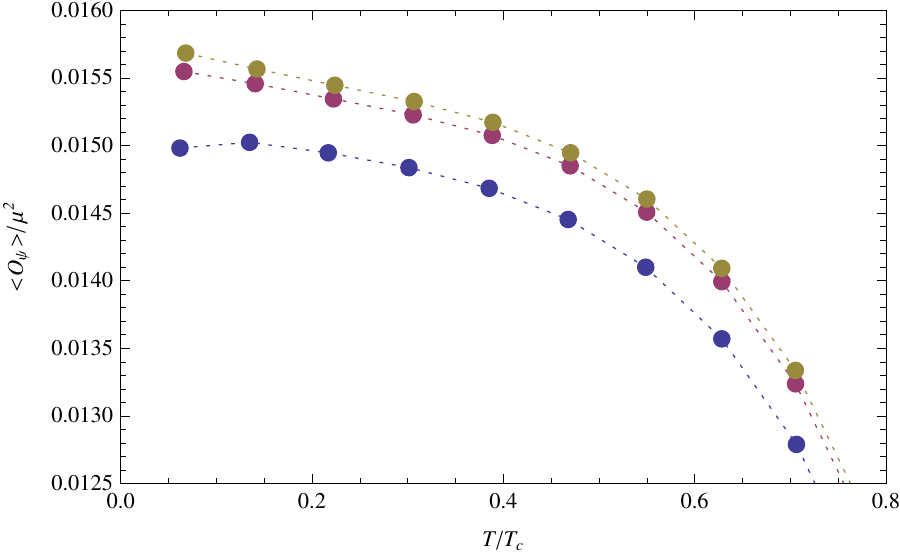}
    \caption[]{The value of the scalar field condensate for varying grid sizes for $c_1=8$ and $L\mu/4=0.75$. From top to bottom, the grid spacing is $d\rho, dx={0.01,0.02,0.04}$. The results are consistent with second order scaling as expected from our numerical approach.}
\label{Opsi_cnvrg}
\end{figure}

The UV procedure is unstable for a generic initial guess, resulting in a divergent norm. To find a solution from a generic initial guess, we can run the relaxation without the UV procedure until the norm is small enough that the result approximates the true solution, before activating the UV procedure to find the true solution. Once we have these first solutions, by using these as an initial guess for solutions nearby in parameter space and by interpolating to a finer grid, we can generate further solutions by relaxing with the UV procedure. 
In Fig.~\ref{resids}, we plot the $L^2$ norm of the total residual during the relaxation of the $c_1=8$ solution at $T_0=0.04$ and $L\mu/4=0.75$ for the grid spacings $d\rho, dx={0.04,0.02,0.01}$, showing the expected exponential behavior of the Gauss-Siedel relaxation. The physical data extracted from our solutions is consistent with the expected second order convergence of our FDA scheme, see Fig.~\ref{Opsi_cnvrg}.

\subsubsection{Asymptotic versus first law mass}
A useful check of the numerics is to compare the mass of the system read off from the asymptotics of the metric, equation (\ref{mass}), to that computed by integrating the first law, equation (\ref{first}). Since the temperature and entropy are read off from the horizon, comparing these two methods of finding the mass provides a non-trivial global consistency check on our results. We verify that the difference between the asymptotic mass and the first law mass remains smaller than 0.5\% across our set of trials, indicating consistency of our results.

A related check of the numerics is the conformal identity or the Smarr-like relation, $2M=TS+\mu N -\tau_x L$, derived above from the first law for the finite length system. To evaluate how well our solutions satisfy this equation, we examine the ratio \be \frac{2M_{fall-off}-TS-\mu N +\tau_x L}{\textrm{max}(M_{fall-off},TS,\mu N, \tau_x L)}, \ee since the largest term in the expression sets a scale for the cancellation we expect. This ratio is very small for our solutions near the critical temperature. As we lower the temperature, this ratio increases, but stays small. The precise value depends on the parameters of the solution, but is not larger than order 1\%. Moreover, this ratio decreases as we move the position of the finite cutoff of the conformal boundary to a larger radius.

\subsubsection{Finite $\rho_{cut}$ boundary check}
For the $c_1=8$ trials reported in the paper, we use $\rho_{cut}=12$ as our conformal boundary. In Table \ref{rhocut} we present results for varying $\rho_{cut}$, showing that our choice is large enough such that the physical results are insensitive to the cutoff. Although the physical results presented in the table appear very stable, at small $\rho_{cut}$, the results for the mass and charge depend significantly on the fitting procedure for the asymptotic metric functions and gauge field. By running our simulations at $\rho_{cut}=12$, we are both well within the the region where the solutions do not change with the conformal boundary and within a region where our fitting procedure to the asymptotics behaves well.

\begin{table}[h!]
\centering
\begin{tabular}{| c | c | c | c | c |}
\hline
$\rho_{cut}$ & $S$ & $M$ & $N$ \\
\hline
1 & 0.758504 & 0.305774 & 0.527406\\
2 &0.767913 & 0.342327 & 0.490524\\
3 & 0.768211 & 0.341928 & 0.490593\\
4 & 0.768285 & 0.342043 & 0.490583\\
5 & 0.768311 & 0.342136 & 0.490577\\
6 & 0.768322 & 0.34221 & 0.490574\\
7 & 0.768328 & 0.342277 & 0.490572\\
8 & 0.768332 & 0.342324 & 0.49057\\
9 & 0.768334 & 0.342367 & 0.490569\\
10 & 0.768335 & 0.342402 & 0.490568\\
11 & 0.768336 & 0.342434 & 0.490568\\
12 & 0.768336 & 0.342459 & 0.490567\\
\hline
\end{tabular}
\caption{Behavior of physical quantities with the cutoff for $c_1=8$ and $L\mu/4=0.75$ and for fixed grid resolution $d\rho,dx\sim0.02$. The entropy $S$ is read off at the horizon, while the mass $M$ and the charge $N$ are read off at the conformal boundary. Both the entropy and the charge are very robust against the location of the conformal boundary. The mass takes slightly longer to settle down, but is well within the convergent range for $\rho_{cut}=12$.}
\label{rhocut}
\end{table}

\subsubsection{Behavior of the constraints}

\begin{figure}
\center
    \includegraphics[width=12cm]{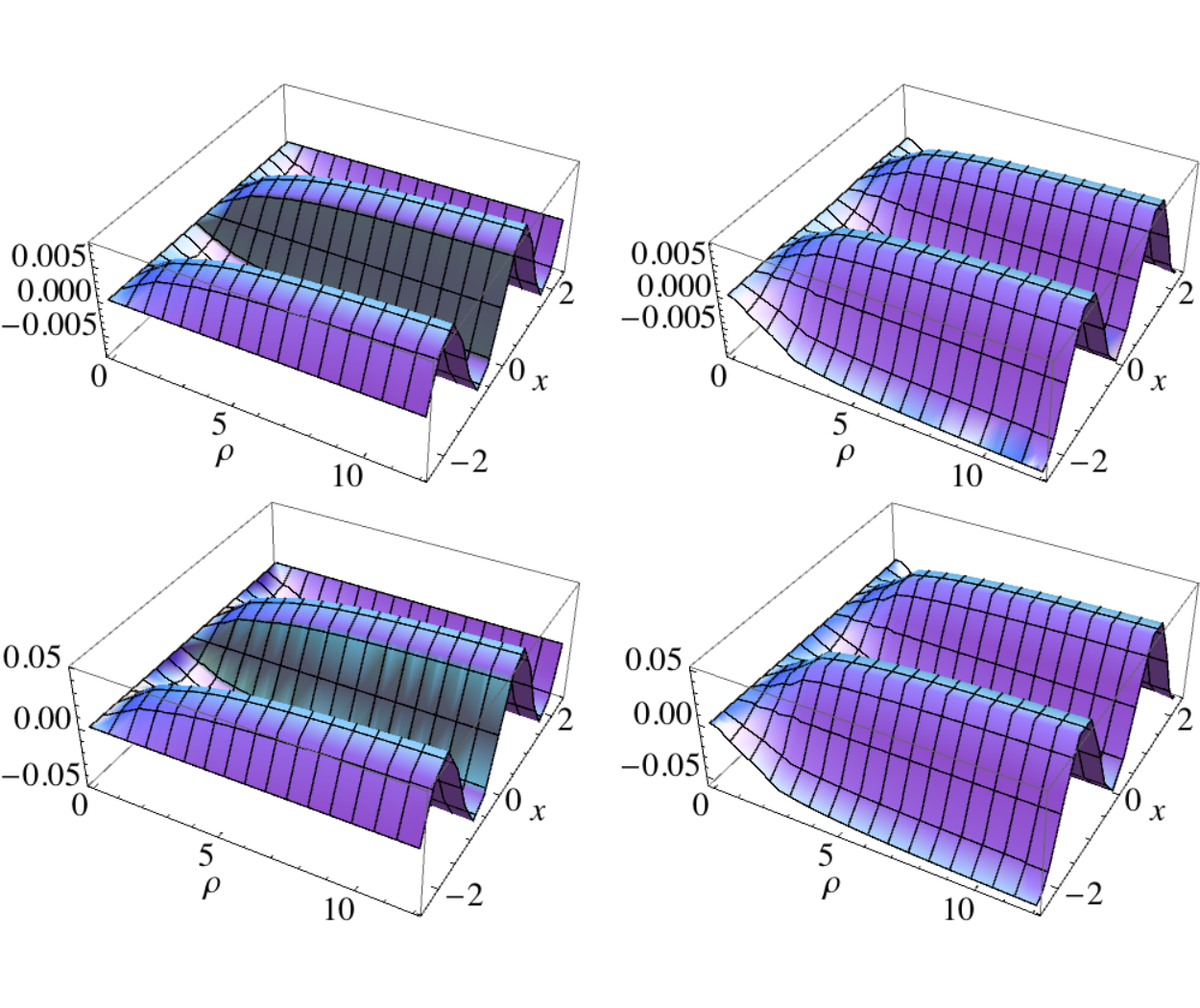}
    \put(-70,276){\makebox(0,0){$r^2\sqrt{f}\sqrt{-g}(G^r_r-G^x_x-(T^r_r-T^x_x))$}}
    \put(-260,276){\makebox(0,0){$\sqrt{-g}(G^r_x-T^r_x)$}}
    \caption[]{The weighted constraints for $c_1=8$ and $L\mu/4=1.21$. The top plots are near the critical point, $T/T_c=0.97$, while the bottom plots are at small temperature, $T/T_c=0.00016$. By our boundary conditions, the constraints disappear at the horizon. They approach a finite value as they approach the asymptotic boundary.}
\label{cons}
\end{figure}

One of the most important checks for our numerical solution is the behavior of the constraints. For numerical homogeneous solutions found with our method, the $L^2$ norm of the constraints is very small, on the order of $10^{-4}$. For the inhomogeneous solutions, the constraints are small near the critical temperature, but grow and saturate as we lower to the temperature, to have a maximum $L^2$ norm on the order of $10^{-2}$: see Fig.~\ref{cons}. Since our boundary conditions explicitly fix the weighted constraints on the horizon, they disappear there. The weighted constraints then increase towards the conformal boundary, approaching a modulated profile of constant amplitude. The amplitude near the conformal boundary controls the overall $L^2$ norm of the constraints. 

The constraint violation improves marginally with step size and with moving the interface closer to the horizon, but does not improve as we take the conformal boundary to a larger radius. To check that the constraints are well satisfied on our solution, we compare them to the sum of the absolute value of the terms that make up the constraints. That is, if the constraints are given by $\sum_i h_i$, we compare this to $\sum_i |h_i|$. This procedure gives us an idea of the scale of the cancellation among the individual terms $h_i$. We find that the sum $\sum_i |h_i|$ diverges approximately as $r^4$ towards the asymptotic boundary, such that the approach of the constraint violation to a constant is a good indicator that the constraints are satisfied on the solution. In Table \ref{con_tab}, we compare the $L^2$ norm of these two sums on the entire domain, showing that the constraint violation for the inhomogeneous solutions is generally about four orders of magnitude less than the scale set by $\sum_i |h_i|$. Interestingly, the relative constraint improves marginally as we go to lower temperatures.

\begin{table}[h!]
\centering
\begin{tabular}{| l | c | c |}
\hline
Parameters & $T_0$ & $L^2(\sum_i h_i)/L^2(\sum_i |h_i|)$ \\
\hline
$c_1=8,L\mu/4=2.00$ (RN solution) & 0.105 & $9.12\cdot10^{-7}$ \\
$c_1=8,L\mu/4=1.21$ (striped solution) & 0.075 & $2.02\cdot10^{-4}$  \\
 & 0.05 & $1.84\cdot10^{-4}$ \\
 & 0.025 & $1.58\cdot10^{-4}$ \\
 & 0.005 & $1.37\cdot10^{-4}$  \\
 & 0.001 & $1.32\cdot10^{-4}$ \\
 \hline
\end{tabular}
\caption{Comparison of the constraint violation, measured by the schematic constraint equation $\sum_i h_i$, to the scale set by the individual terms, $\sum_i |h_i|$, for grid size $d\rho,dx\sim0.01$. We take the $L^2$ norm of the measures on the entire domain. The $c_1=8,L\mu/4=2.00$ solution is a homogeneous RN solution found numerically with our code, for which the constraints are very well satisfied. The constraints for the striped solutions are satisfied compared to the scale set by $\sum_i |h_i|$ by four orders of magnitude and the relative constraint improves marginally as we lower the temperature.}
\label{con_tab}
\end{table}

\subsubsection{The asymptotic equation of motion}

Expanding the equations of motion asymptotically gives the relation \be R^{(3)}(x)+2S^{(3)}(x)+T^{(3)}(x)=0,\label{eomcheck}\ee which can be used to give another check of the numerics. As explained in \ref{appx_inhomo_charges}, this condition implies the tracelessness of the energy-momentum tensor. For the inhomogeneous solutions near the critical temperature we find that this expression is on the order of the individual metric functions $X^{(3)}$, where $X=\{R,S,T\}$, but generally decreases as we lower the temperature. As well, we find that homogeneous solutions found using our numerical techniques satisfy (\ref{eomcheck}) well. There seems to be an unidentified systematic error here that may deserve further attention in the future. Possible problems may occur in the implementation of the UV procedure or in our procedure to read off the coefficients of the falloffs of the metric functions. However, our physical results are robust under changes to the boundary conditions, so that we are confident in our results despite this possible systematic. In particular, the physical quantities extracted from the horizon are independent of the different boundary constraint fixing schemes we implemented. Therefore, we advocate using the mass derived from the integrated first law, which uses no asymptotic metric functions.

%====================================================================%

%====================================================================%
\end{document}